\documentclass[acmsmall, nonacm]{acmart}

\usepackage[textsize=scriptsize,textwidth=3em]{todonotes}

\usepackage{graphicx}
\usepackage{pifont}
\usepackage{hyperref}

\newcommand{\cm}{\ding{51}}%
\newcommand{\xm}{\ding{55}}%

\usepackage{xcolor}
\usepackage[normalem]{ulem} % For strikethrough
\usepackage{changepage}
\usepackage{caption}
\AtBeginDocument{%
  \providecommand\BibTeX{{%
    \normalfont B\kern-0.5em{\scshape i\kern-0.25em b}\kern-0.8em\TeX}}}

\setcopyright{none}
\renewcommand\footnotetextcopyrightpermission[1]{} % removes footnote with conference information in first column
\graphicspath{{./images/}}

\begin{document}

% \tableofcontents
% \newpage

%%
%% The "title" command has an optional parameter,
%% allowing the author to define a "short title" to be used in page headers.
\title{Memory Under Siege: A Comprehensive Survey of Side-Channel Attacks on Memory}

\author{Md Mahady Hassan}
\email{mdhassan@augusta.edu}
\affiliation{%
  \institution{Augusta University}
  \city{Augusta}
  \state{Georgia}
  \postcode{30904}
  \country{USA}
}

% \author{Md Ashiqur Rahman}
% \email{mdrahman@augusta.edu}
% \affiliation{%
%   \institution{Augusta University}
%   \city{Augusta}
%   \state{Georgia}
%   \postcode{30904}
%   \country{USA}
% }

\author{Shanto Roy}
\email{sroy10@uh.edu}
\affiliation{%
  \institution{University of Houston}
  \city{Houston}
  \state{Texas}
  \postcode{77204}
  \country{USA}
}

\author{Reza Rahaeimehr}
\email{rrahaeimehr@augusta.edu}
\affiliation{%
  \institution{Augusta University}
  \city{Augusta}
  \state{Georgia}
  \postcode{30904}
  \country{USA}
}

%%
%% The "author" command and its associated commands are used to define
%% the authors and their affiliations.
%% Of note is the shared affiliation of the first two authors, and the
%% "authornote" and "authornotemark" commands
%% used to denote shared contribution to the research.

%%
%% By default, the full list of authors will be used in the page
%% headers. Often, this list is too long, and will overlap
%% other information printed in the page headers. This command allows
%% the author to define a more concise list
%% of authors' names for this purpose.
\renewcommand{\shortauthors}{M.\ Hassan et al.}

%%
%% The abstract is a short summary of the work to be presented in the
%% article.
\begin{abstract}
Side-channel attacks on memory (SCAM) exploit unintended data leaks from memory subsystems to infer sensitive information, posing significant threats to system security. These attacks exploit vulnerabilities in memory access patterns, cache behaviors, and other microarchitectural features to bypass traditional security measures. The purpose of this research is to examine SCAM, classify various attack techniques, and evaluate existing defense mechanisms. It guides researchers and industry professionals in improving memory security and mitigating emerging threats. We begin by identifying the major vulnerabilities in the memory system that are frequently exploited in SCAM, such as cache timing, speculative execution, \textit{Rowhammer}, and other sophisticated approaches. Next, we outline a comprehensive taxonomy that systematically classifies these attacks based on their types, target systems, attack vectors, and adversarial capabilities required to execute them. In addition, we review the current landscape of mitigation strategies, emphasizing their strengths and limitations. This work aims to provide a comprehensive overview of memory-based side-channel attacks with the goal of providing significant insights for researchers and practitioners to better understand, detect, and mitigate SCAM risks.
\end{abstract}

%%
%% The code below is generated by the tool at: http://dl.acm.org/ccs.cfm
%% Please copy and paste the code instead of the example below.
%%
% \begin{CCSXML}
% <ccs2012>
%  <concept>
%   <concept_id>00000000.0000000.0000000</concept_id>
%   <concept_desc>Do Not Use This Code, Generate the Correct Terms for Your Paper</concept_desc>
%   <concept_significance>500</concept_significance>
%  </concept>
%  <concept>
%   <concept_id>00000000.00000000.00000000</concept_id>
%   <concept_desc>Do Not Use This Code, Generate the Correct Terms for Your Paper</concept_desc>
%   <concept_significance>300</concept_significance>
%  </concept>
%  <concept>
%   <concept_id>00000000.00000000.00000000</concept_id>
%   <concept_desc>Do Not Use This Code, Generate the Correct Terms for Your Paper</concept_desc>
%   <concept_significance>100</concept_significance>
%  </concept>
%  <concept>
%   <concept_id>00000000.00000000.00000000</concept_id>
%   <concept_desc>Do Not Use This Code, Generate the Correct Terms for Your Paper</concept_desc>
%   <concept_significance>100</concept_significance>
%  </concept>
% </ccs2012>
% \end{CCSXML}

% \ccsdesc[500]{Do Not Use This Code~Generate the Correct Terms for Your Paper}
% \ccsdesc[300]{Do Not Use This Code~Generate the Correct Terms for Your Paper}
% \ccsdesc{Do Not Use This Code~Generate the Correct Terms for Your Paper}
% \ccsdesc[100]{Do Not Use This Code~Generate the Correct Terms for Your Paper}

%%
%% Keywords. The author(s) should pick words that accurately describe
%% the work being presented. Separate the keywords with commas.
\keywords{Side-channel attacks, Memory side channels, Cache attacks, Timing side channels, Information leakage, Microarchitectural vulnerabilities}

%% The following are not a requirement, delete if not using
% \received{20 February 2024}  %% inital submission date
% \received[revised]{12 March 2024} %% interim new draft
% \received[accepted]{5 June 2024}  %% publication version

%%
%% This command processes the author and affiliation and title
%% information and builds the first part of the formatted document.
\settopmatter{printfolios=true}
\maketitle

\section{Introduction}
% \TODO{Significance of this work}

In recent years, the proliferation of computing devices and the increasing complexity of software systems have led to a significant rise in the sophistication of cyber attacks~\cite{roy2022survey}. Side-channel attacks on memory (SCAM) have emerged as a critical threat, exploiting unintentional data leakage from a system's memory during its operation~\cite{lou2021survey,lyu2018survey,anwar2017cross,wang2023side,su2021survey,baviskar2024cache}. These attacks can reveal sensitive data, such as cryptographic keys~\cite{banegas2023recovering, dziembowski2008leakage, goubin2002refined, wang2023nvleak, li2021cipherleaks, lou2021survey, li2021cipherleaks, dong2018shielding}, user credentials~\cite{kwong2023checking, wang2017side, naghibijouybari2018rendered, naghibijouybari2019side, unterluggauer2016exploiting, lipp2021platypus, wang2023nvleak}, and other confidential information, compromising the security of various applications, including cloud computing, mobile devices, and embedded systems~\cite{alahmadi2022cyber, lyu2018survey, bazm2017side, mendez2021physical, tsoupidi2023thwarting, fournaris2017exploiting, abrishamchi2017side, karimi2018timing, ahmed2024deep}.

% \TODO{Research Gap}
Although several studies have focused on specific types of side-channel attacks, such as cache timing attacks and \textit{Rowhammer} exploitation ~\cite{gruss2016flush+, yarom2014flush+, shusterman2021prime+, disselkoen2017prime+, younis2015new, cheng2024evict+, mutlu2019rowhammer, van2016drammer, gruss2016rowhammer}, there is a lack of a unified framework that categorizes these attacks and elucidates their underlying principles. Furthermore, the dynamic nature of memory architectures and the continuous evolution of attack techniques require an updated survey that reflects the current state of research and practice in this domain. To address these gaps, this survey paper aims to answer the following research questions

\vspace{0.5em}
{\small 
\begin{adjustwidth}{3.5em}{1em}
\begin{itemize}
    \item[\textbf{RQ1.}] \textit{What are the viable targets for SCAM, and why are these targets attractive to adversaries?}
    \item[\textbf{RQ2.}] \textit{What are the various types of SCAM, and how to categorize these attacks?}
    \item[\textbf{RQ3.}] \textit{What countermeasures have been proposed to mitigate the risks associated with these attacks, and how effective are they in practice?}
\end{itemize}
\end{adjustwidth}
}
% \TODO{Answers/Contributions}

%% Rewrite by Ashiq
In response to these research questions, we first examine the attack surface exploited in SCAM, providing insight into why these targets are particularly vulnerable to adversaries (RQ1). 
We then present a comprehensive taxonomy of SCAM, categorizing various types of side-channel attacks and analyzing their operational mechanisms, techniques, and associated tools (RQ2). In addition, we evaluate existing countermeasures to assess their effectiveness, identify limitations, and propose future research directions to enhance the resilience of the system against these threats (RQ3). This survey aims to equip researchers and practitioners with a deeper understanding of SCAM, offering a structured overview that informs both defensive strategies and future advancements in cybersecurity.

%%In response to these research questions, we answer the RQ1 by looking into the attack surface used for SCAMs. Here, we identify why these targets are often exploited by adversaries. Then we answer RQ2 by presenting a comprehensive taxonomy of SCAMs where we categorize different types of side-channel attacks. To address the second question, we look into the mechanisms of these attacks and then investigate further on their operational characteristics, techniques, and used tools. Furthermore, we analyze existing countermeasures to answer RQ3, highlight their strengths and weaknesses, and propose future research directions to enhance systems' resilience against these pervasive threats. This survey aims to equip researchers and practitioners with the knowledge to better understand and defend against these sophisticated attack vectors by providing a detailed overview of SCAMs.

%%Contribution by Ashiq
%This survey contributes to the field of cybersecurity by providing a structured and current analysis of SCAMs. Identifies key targets of these attacks and explains their susceptibility to exploitation. The paper introduces a taxonomy that categorizes SCAMs on the basis of their techniques and operational mechanisms. In addition, it evaluates existing countermeasures, highlighting their limitations, and proposing directions for future research. These efforts offer a concise overview that supports the development of more effective defensive strategies.

% \TODO{Finalize Organization at the end...}

The rest of this paper is structured as follows: Section~\ref{sec:background}  conveys the background knowledge of various types of attacks. Section \ref{sec:taxonomy} introduces a novel taxonomy to present and categorize various attacks. Section \ref{sec:side-channel-attack-details} describes the attack methodologies in detail and includes five comparison tables, each corresponding to a broad category of attacks. Section~\ref{sec:lesson} discusses various mitigation strategies against these attacks and presents a comparative analysis of the proposed techniques. Finally, Section~\ref{sec:conclusion} summarizes the key points of this systematization work, including suggestions for future works.

\section{Background} 
\label{sec:background}

SCAM exploits unintended data leaks in memory subsystems, such as caches, DRAM, and page tables, to infer sensitive data without directly targeting cryptographic algorithms or software vulnerabilities~\cite{gruss2016flush+, yarom2014flush+, shusterman2021prime+, disselkoen2017prime+, cheng2024evict+, mutlu2019rowhammer, van2016drammer, wang2017leaky}. As a specialized subset of side-channel attacks, including techniques exploiting power consumption and electromagnetic emissions~\cite{kocher1999differential}, memory side-channels have gained prominence due to the growing complexity of modern computing systems. 

First introduced by \textit{Kocher et al.}~\cite{kocher1996timing} through timing-based techniques on cryptographic systems, SCAM has significantly evolved, leveraging cache timing variations~\cite{bernstein2005cache}, shared memory exploits~\cite{yarom2014flush+, gruss2016flush+, shusterman2021prime+, cheng2024evict+}, and physical vulnerabilities like \textit{Rowhammer}~\cite{mutlu2019rowhammer}. Techniques such as \textit{Flush+Reload} and \textit{Prime+Probe} exploit timing differences in shared caches~\cite{yarom2014flush+, osvik2006cache}, while \textit{Rowhammer} targets DRAM vulnerabilities to induce bit flips~\cite{kim2014flipping, van2016drammer}. The rise of speculative execution has further expanded the attack surface, as demonstrated by \textit{Spectre} and \textit{Meltdown}~\cite{kocher2020spectre, lipp2020meltdown}. These attacks have implications across various domains, from multi-tenant cloud environments to embedded systems and IoT devices~\cite{tsoupidi2023thwarting, fournaris2017exploiting, abrishamchi2017side, zhang2012cross, alahmadi2022cyber, bazm2017side}, where resource sharing increases the potential for data leakage. As memory architectures become more complex, SCAM represents a significant and evolving threat to data security in critical systems.

\subsection{CPU and Memory Targets} 
Building on the basics of SCAM, we now focus on the architectural features of modern CPUs and memory systems that make them vulnerable to exploitation. Designed for high-speed computation, components such as branch target buffers (BTB)~\cite{evtyushkin2018branchscope, kong2012architecting}, return stack buffers (RSB)~\cite{koruyeh2018spectre, kim2019high}, indirect branch predictors (IBP)~\cite{chowdhuryy2021leaking}, speculative execution mechanisms~\cite{kocher2020spectre}, memory layouts like caches~\cite{yarom2014flush+, osvik2006cache}, translation lookaside buffers (TLB)~\cite{agredo2024inferring, gras2018translation}, DRAM row buffers~\cite{mutlu2019rowhammer}, and page tables~\cite{koschel2020tagbleed} often introduce unintended pathways for data leakage. Although these features are essential for maximising performance, they expose systems to sophisticated attacks such as \textit{Spectre}, \textit{Meltdown}, \textit{Flush+Reload}, \textit{Prime+Probe}, \textit{Rowhammer}~\cite{kocher2020spectre, lipp2020meltdown,yarom2014flush+, osvik2006cache, kim2014flipping}, etc. By understanding how these features interact with adversarial techniques, we can better grasp the balance between performance optimisation and security risks in modern computing.

\subsubsection{\textbf{Branch Target Buffers (BTB)}}
Branch target buffers are small cache-like structures in modern CPUs that store the predicted target addresses of branch instructions, allowing for faster execution~\cite{chowdhuryy2020branchspec}. They play a critical role in speculative execution by allowing the CPU to execute instructions based on branch predictions~\cite{kocher2020spectre, canella2019systematic}. However, attackers exploit BTB in speculative execution attacks like \textit{Spectre Variant 2}~\cite{kocher2020spectre, koruyeh2018spectre} by training the buffer to mispredict the logic in advance, causing unauthorised instructions to leave measurable side-channel traces.

\subsubsection{\textbf{Return Stack Buffers (RSB)}}
Return stack buffers are small structures in modern CPUs that predict return addresses for function calls, allowing faster execution by avoiding delays in return address resolution~\cite{maisuradze2018ret2spec}. These buffers are integral to speculative execution, enabling the CPU to maintain performance during complex branching~\cite{wikner2022retbleed}. In speculative execution attacks such as \textit{SpectreRSB}, adversaries exploit RSB by overwriting buffer entries with malicious return addresses~\cite{olmos2024protecting}. This misdirection causes the CPU to speculatively execute unintended instructions, leaving cache traces that attackers can analyze to extract sensitive data~\cite{koruyeh2018spectre}.

\subsubsection{\textbf{Indirect Branch Predictors (IBP)}}
Indirect branch predictors are components in modern CPUs designed to predict the targets of indirect branches, enabling the CPU to execute instructions speculatively for improved performance~\cite{wikner2024breaking}. By storing historical branch information, these predictors help reduce the delays associated with indirect branch resolution~\cite{maisuradze2018ret2spec, kimslap}. Their role in speculative execution makes them an attractive attack target, as adversaries can manipulate prediction patterns to influence speculative execution paths~\cite{kocher2020spectre}.

\subsubsection{\textbf{Speculative and Out-of-Order Execution}} Speculative and out-of-order execution are core mechanisms in modern CPUs that enhance performance by executing instructions ahead of branch resolution or in a non-sequential order. These optimizations allow the CPU to maximize resource utilization and reduce execution stalls, significantly improving throughput in complex applications. Despite their advantages, these mechanisms introduce critical vulnerabilities. Speculative execution has been exploited in attacks like \textit{Spectre} and \textit{Meltdown}, where speculative operations on unauthorized data leave observable side-channel traces in shared resources like caches~\cite{kocher2020spectre, lipp2020meltdown}. Attackers can analyze these traces to extract sensitive information, such as encryption keys or passwords~\cite{maisuradze2018ret2spec, kwong2023checking, mcilroy2019spectre}. Mitigations such as speculative execution barriers and a stronger process isolation technique have been proposed to address these risks~\cite{intel_retpoline_2018, kiriansky2018speculative}. However, these defenses often come with performance trade-offs, underscoring the challenge of balancing security with efficiency in modern CPU design.

\subsubsection{\textbf{Prefetching}} Prefetching is a CPU optimization that predicts and loads data or instructions into the cache before they are explicitly requested, reducing latency and improving performance. However, this mechanism can unintentionally leak memory access patterns, making it a target for side-channel attacks~\cite{gruss2016prefetch, irazoqui2017cross}. By exploiting these patterns, attackers can bypass defenses like Address Space Layout Randomization (ASLR) and infer sensitive information~\cite{hund2013practical, gras2017aslr, lipp2022amd}.

\subsubsection{\textbf{Hardware Transactional Memory (HTM)}} Hardware Transactional Memory allows the atomic execution of instruction groups, improving concurrency and minimizing the overhead of traditional locking mechanisms~\cite{gruss2017strong, shih2017t}. While beneficial for performance, HTM introduces vulnerabilities to timing-based side-channel attacks. For example, transaction failures can leak sensitive data through precise timing measurements, allowing attackers to infer memory access patterns~\cite{chen2019exploitable, ferracci2019detecting}. Techniques like \textit{Prime+Abort} exploited Intel TSX, where abort timings reveal memory access patterns~\cite{disselkoen2017prime+, guan2015protecting}.

\subsubsection{\textbf{Translation Lookaside Buffers (TLB) and Page Tables}} Translation Lookaside Buffers are small, fast caches in modern CPUs that store recently used Virtual-to-Physical address mappings, improving memory access speed by reducing Page Table lookups~\cite{van2017revanc}. Page Tables maintain a hierarchical map of all Virtual-to-Physical translations, enabling efficient memory management in complex systems~\cite{strackx2017heisenberg}. TLB and Page Tables can leak sensitive information through timing variations~\cite{gras2018translation}. For instance, attackers can exploit page fault timing or TLB misses to infer memory access patterns, bypassing defenses like ASLR \cite{koschel2020tagbleed, gruss2016prefetch}.

\subsubsection{\textbf{Cache Hierarchies (L1, L2, L3)}} 
Modern CPUs feature a multilevel cache hierarchy, including L1, L2, and L3 caches, to bridge the speed gap between the CPU and main memory. These caches store frequently accessed data to minimize latency, with L1 being the smallest and fastest, located closest to the CPU core, and L3 being the largest and slowest, shared among multiple cores \cite{lyu2018survey}. By organizing caches in this way, CPUs optimize data retrieval, enhance data locality, and significantly reduce memory access times. Despite these benefits, cache hierarchies are vulnerable to timing-based side-channel attacks, such as \textit{Flush+Reload} and \textit{Prime+Probe}, which exploit variations in access times to infer memory access patterns~\cite{yarom2014flush+, osvik2006cache, gruss2016flush+, irazoqui2017cross}. These attacks are particularly effective in shared environments, such as cloud systems, where different users access the same cache resources, allowing attackers to extract sensitive information such as cryptographic keys or activity patterns~\cite{zhang2012cross, zhao2024you, zhang2014cache}.

\subsubsection{\textbf{DRAM Row Buffers}} In Dynamic Random-Access Memory (DRAM), row buffers are used to temporarily hold data from memory rows before it is read or written~\cite{kim2020revisiting}. When a memory row is accessed, its contents are loaded into the row buffer, significantly reducing the access latency for subsequent reads or writes to the same row~\cite{zhang2022implicit}. This mechanism improves memory performance by exploiting spatial locality~\cite{jacob2010memory}. Despite its benefits, the row buffer behavior can be exploited in side-channel attacks. For example, the \textit{Rowhammer} attack manipulates the row buffer by repeatedly accessing adjacent memory rows, causing bit flips in nearby rows due to electrical interference\cite{gruss2018another, van2018guardion}. Bit flips enable attackers to alter the memory contents, bypassing privilege boundaries and compromising data integrity~\cite{kim2014flipping, aga2017good}.

\subsubsection{\textbf{Shared Memory Resources}} Shared memory resources, such as DRAM and shared caches, increase computational speed and minimize memory overhead in multi-core and multi-tenant systems~\cite{vano2018slicedup}. In virtualized environments, memory deduplication further enhances efficiency by consolidating identical memory pages~\cite{barham2003xen, maurice2015information}. These benefits come with security risks. Memory deduplication can be exploited in attacks, where timing variations reveal shared pages and leak sensitive information~\cite{zhao2024you, bosman2016dedup}. Resource sharing in caches and DRAM can reveal memory access patterns, enabling side-channel attacks in multi-tenant environments like cloud platforms~\cite{zhang2012cross}.

\subsection{Common Attack Models} Side-channel attacks on memory can be classified according to the proximity and activity of the attacker. Local attacks require the adversary to have physical or low-level access to the target system, while remote attacks exploit vulnerabilities over a network without direct access. Additionally, attackers may operate passively, observing traces such as timing variations and power usage, or actively, by deliberately inducing faults or triggering unexpected behaviors through techniques like voltage or frequency manipulation. These classifications underscore the diverse strategies adversaries employ to compromise modern computing systems.
\section{Proposed Taxonomy} \label{sec:taxonomy}
\begin{figure*}[!ht]
    \centering
    \includegraphics[width=\textwidth]{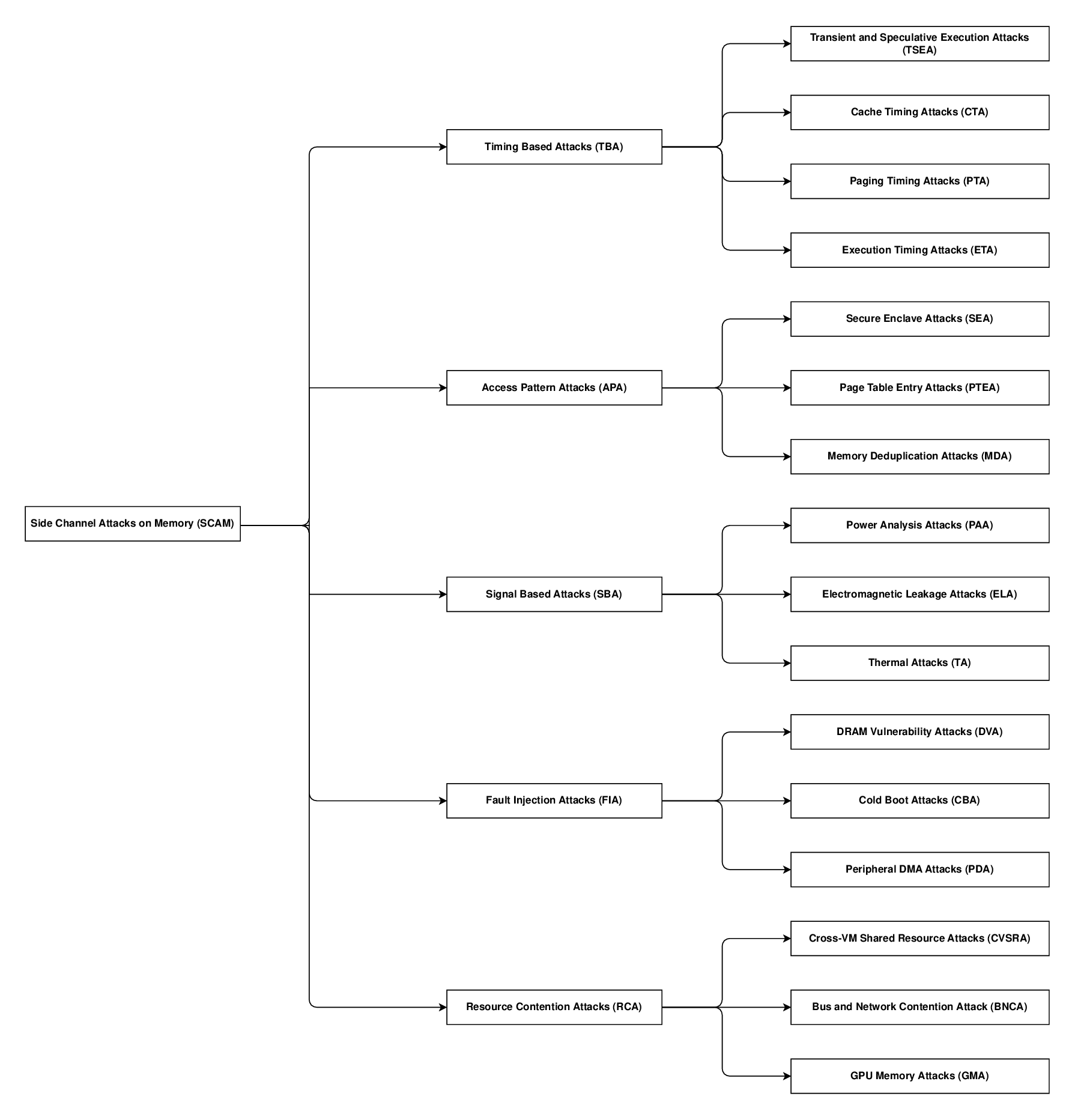}
    \caption{SCAM Taxonomy}
    \label{fig:recon_taxonomy_info}
\end{figure*}

SCAM exploits unintended information leaks from memory access patterns, timing variations, signal emissions, or resource contention to infer sensitive data~\cite{lou2021survey, jiang2020mempoline, ali2023characterization}. To comprehensively categorize these threats, we propose a taxonomy that classifies side-channel attacks on memory into five primary categories: Timing-Based Attacks (TBA), Access Pattern Attacks (APA), Signal-Based Attacks (SBA), Fault Injection Attacks (FIA), and Resource Contention Attacks (RCA). Each category encapsulates unique attack methodologies while considering different leakage sources and targeting different hardware components. This taxonomy enables a structured understanding of the attack landscape and facilitates a comparative analysis of existing works. Furthermore, this taxonomy can guide future research towards comprehensive mitigation methods. Figure~\ref{fig:recon_taxonomy_info} presents five primary attack categories, along with examples that fall under each.

\section{Side Channel Attacks on Memory (SCAM)} \label{sec:side-channel-attack-details}

This section expands upon the SCAM taxonomy introduced earlier by systematically analyzing each of the five primary categories of side-channel attacks on memory, along with their subcategories. The goal is to provide a detailed understanding of the underlying mechanisms, attack surfaces and practical implications. Each subsection discusses attack strategies, leakage sources and hardware targets, illustrating how adversaries exploit memory behaviors to infer sensitive information.

\subsection{Timing-Based Attacks (TBA)}
Timing-based attacks exploit execution timing variations to extract sensitive data, such as cryptographic keys or memory contents. By analyzing these variations, attackers can infer system states or operations, exposing vulnerabilities inherent in modern computing architectures. These attacks are particularly effective due to their ability to operate without direct access to the target system, relying solely on measurable timing information.
TBA can be categorized into four subtypes based on the distinct microarchitectural behaviors they exploit. These subcategories include transient and speculative execution attacks (TSEA), cache timing attacks (CTA), paging timing attacks (PTA), and execution timing attacks (ETA), each targeting distinct microarchitectural behaviors. Table~\ref{tab:comparison-table-tba} presents a comparison of existing literature that focused on timing-based SCAM.

TSEA exploits vulnerabilities in speculative CPU operations, enabling attackers to access sensitive data through transient instructions before they are committed~\cite{kocher2020spectre, lipp2020meltdown}. Similarly, CTA analyzes access time discrepancies in cache memory, using methods like \textit{Prime+Prob}e and \textit{Flush+Reload}~\cite{liu2015last, yarom2014flush+} to deduce sensitive information. PTA leverages timing variations in memory management mechanisms, such as translation lookaside buffer and page faults~\cite{gras2018translation, wang2017leaky}. ETA exploits broader timing variations in system operations, such as those observed during decompression or in high-resolution timing analysis~\cite{schwarzl2023practical, schwarz2017fantastic}.

% Please add the following required packages to your document preamble:
% \usepackage{multirow}
\begin{table*}[!ht]
\centering
\caption{Comparison of Different Timing-Based Memory Side-Channel Attacks}
\label{tab:comparison-table-tba}
\resizebox{\textwidth}{!}{
\begin{tabular}{|c|c|c|c|c|c|c|c|c|c|c|}
\hline
\multirow{2}{*}{\textbf{Literature (Year)}} &
\multirow{2}{*}{\begin{tabular}[c]{@{}c@{}}\textbf{Attack}\\\textbf{Type}\end{tabular}} &
\multirow{2}{*}{\textbf{Methodology}} &
\multicolumn{4}{c|}{\textbf{Platform}} &
\multirow{2}{*}{\textbf{Target}} &
\multirow{2}{*}{\textbf{Impact}} &
\multicolumn{2}{c|}{\textbf{Mitigation}} \\ \cline{4-7} \cline{10-11}
& & & \textbf{Intel} & \textbf{AMD} & \textbf{ARM} & \textbf{Cross} & & & \textbf{HW} & \textbf{SW} \\ \hline\hline

Bhattacharyya et al. (2019) &
TSEA &
Port Contention &
\cm &  % Intel
\xm &  % AMD
\xm &  % ARM
\xm &  % Cross
Execution Ports &
Key Leakage &
\xm &  % HW Mitigation
\xm \\ \hline  % SW Mitigation

Gruss et al. (2015) &
TSEA, CTA &
Flush+Reload &
\cm &  % Intel
\cm &  % AMD
\cm &  % ARM
\xm &  % Cross
L3 Cache &
Key Leakage &
\xm &  % Hw Mitigation
\cm \\ \hline  % SW Mitigation

Li et al. (2024) &
TSEA, CTA &
Indirect Branch Injection &
\cm &  % Intel
\xm &  % AMD
\xm &  % ARM
\xm &  % Cross
Indirect Branch Predictor &
ASLR Bypass &
\xm &  % HW Mitigation
\cm \\ \hline  % SW Mitigation

Lipp et al. (2020) &
TSEA, CTA &
Speculative Execution &
\cm &  % Intel
\xm &  % AMD
\xm &  % ARM
\xm &  % Cross
Kernel Memory &
Privilege Bypass &
\cm &  % HW
\cm \\ \hline  % SW

Kocher et al. (2020) &
TSEA, CTA &
Speculative Execution &
\cm &  % Intel
\cm &  % AMD
\cm &  % ARM
\cm &  % Cross
User space Memory &
Privilege Bypass &
\cm &  % HW
\cm \\ \hline  % SW

Koruyeh et al. (2018) &
TSEA, CTA &
RSB-based Speculation &
\cm &  % Intel
\xm &  % AMD
\xm &  % ARM
\xm &  % Cross
Return Stack Buffer &
Privilege Bypass &
\cm &  % HW Mitigation
\cm \\ \hline  % SW Mitigation

Wikner \& Razavi (2022) &
TSEA, CTA &
Return Stack Injection &
\cm &  % Intel
\cm &  % AMD
\xm &  % ARM
\xm &  % Cross
Return Instructions &
Privilege Bypass &
\cm &  % HW Mitigation
\cm \\ \hline  % SW Mitigation

Trujillo et al. (2023) &
TSEA, CTA &
Transient Training &
\xm &  % Intel
\cm &  % AMD
\xm &  % ARM
\xm &  % Cross
Return Stack Buffer &
Privilege Bypass &
\xm &  % HW Mitigation
\xm \\ \hline  % SW Mitigation

Van Bulck et al. (2018) &
TSEA, CTA &
L1 Terminal Fault &
\cm &  % Intel
\xm &  % AMD
\xm &  % ARM
\xm &  % Cross
Enclave Memory &
Key Leakage &
\cm &  % HW Mitigation
\cm \\ \hline  % SW Mitigation

Shen et al. (2024) &
TSEA, CTA &
Runahead Execution &
\cm &  % Intel
\cm &  % AMD
\xm &  % ARM
\xm &  % Cross
Branch Prediction Unit &
Key Leakage &
\xm &  % HW Mitigation
\cm \\ \hline  % SW Mitigation

Canella et al. (2019) &
TSEA, CTA &
Store Buffer Forwarding &
\cm &  % Intel
\xm &  % AMD
\xm &  % ARM
\xm &  % Cross
Store Buffer &
Data Leakage &
\xm &  % HW Mitigation
\xm \\ \hline  % SW Mitigation

Islam et al. (2019) &
TSEA, CTA &
Speculative Load Hazard &
\cm &  % Intel
\xm &  % AMD
\xm &  % ARM
\xm &  % Cross
Speculative Load Buffer &
Address Leakage &
\xm &  % HW Mitigation
\xm \\ \hline  % SW Mitigation

Cheng et al. (2024) &
TSEA, CTA &
Evict+Spec+Time &
\cm &  % Intel
\cm &  % AMD
\xm &  % ARM
\cm &  % Cross
Last-Level Cache &
AES Key Recovery &
\xm &  % HW Mitigation
\cm \\ \hline  % SW Mitigation

Chowdhuryy \& Yao (2021) &
TSEA, CTA &
PHT Manipulation &
\cm &  % Intel
\xm &  % AMD
\xm &  % ARM
\xm &  % Cross
Pattern History Table &
Secret Leakage &
\xm &  % HW Mitigation
\xm \\ \hline  % SW Mitigation

Jin et al. (2023) &
TSEA, ETA &
Transient Branch Timing &
\cm &  % Intel
\xm &  % AMD
\xm &  % ARM
\xm &  % Cross
Conditional Branch Path &
Register-based Leak &
\xm &  % HW Mitigation
\cm \\ \hline  % SW Mitigation

Liu et al. (2015) &
CTA &
Prime+Probe &
\cm &  % Intel
\cm &  % AMD
\xm &  % ARM
\cm &  % Cross
L3 Cache &
Key Leakage &
\xm &  % HW Mitigation
\xm \\ \hline  % SW Mitigation

Yarom \& Falkner (2014) &
CTA &
Flush+Reload &
\cm &  % Intel
\cm &  % AMD
\xm &  % ARM
\xm &  % Cross
L3 Cache &
Key Leakage &
\xm &  % HW Mitigation
\cm  \\ \hline  % SW Mitigation

Zhang et al. (2014) &
CTA &
Prime+Probe &
\cm &  % Intel
\cm &  % AMD
\xm &  % ARM
\cm &  % Cross
Last-Level Cache &
Cross-VM Leakage &
\xm &  % HW Mitigation
\cm \\ \hline  % SW Mitigation

Yan et al. (2020) &
CTA &
Access Pattern Profiling &
\cm &  % Intel
\cm &  % AMD
\xm &  % ARM
\cm &  % Cross
L3 Cache &
DNN Archi. Leak &
\xm &  % HW Mitigation
\cm \\ \hline  % SW Mitigation

Briongos et al. (2020) &
CTA &
Precise Cache Eviction &
\cm &  % Intel
\xm &  % AMD
\xm &  % ARM
\xm &  % Cross
L3 Cache &
Key Leakage &
\xm &  % HW Mitigation
\xm \\ \hline  % SW Mitigation

Yarom et al. (2017) &
CTA, ETA &
Cache Bank Conflict &
\cm &  % Intel
\xm &  % AMD
\xm &  % ARM
\xm &  % Cross
Cache Bank &
Key Leakage &
\xm &  % HW Mitigation
\xm \\ \hline  % SW Mitigation

Shusterman et al. (2021) &
CTA, ETA &
Prime+Probe &
\cm &  % Intel
\cm &  % AMD
\cm &  % ARM
\cm &  % Cross
L3 Cache &
Cross Site Leakage &
\xm &  % HW Mitigation
\cm \\ \hline  % SW Mitigation

Oren et al. (2015) &
CTA, ETA &
Prime+Probe (JS) &
\cm &  % Intel
\cm &  % AMD
\cm &  % ARM
\cm &  % Cross
L3 Cache &
Cross Tab Leakage &
\xm &  % HW Mitigation
\cm \\ \hline  % SW Mitigation

Disselkoen et al. (2017) &
CTA, ETA &
Prime+Abort &
\cm &  % Intel
\xm &  % AMD
\xm &  % ARM
\xm &  % Cross
L3 Cache &
Key Leakage &
\xm &  % HW Mitigation
\cm \\ \hline  % SW Mitigation

Aciiçmez (2007) &
CTA, ETA &
I-cache Timing &
\cm &  % Intel
\xm &  % AMD
\xm &  % ARM
\xm &  % Cross
Instruction Cache &
Key Leakage &
\xm &  % HW Mitigation
\xm \\ \hline  % SW Mitigation

Wang et al. (2017) &
PTA &
Page-Fault Variants &
\cm &  % Intel
\xm &  % AMD
\xm &  % ARM
\xm &  % Cross
Enclave Memory &
Key Leakage &
\xm &  % HW Mitigation
\cm \\ \hline  % SW Mitigation

Gruss et al. (2016) &
PTA &
Prefetch Probing &
\cm &  % Intel
\xm &  % AMD
\cm &  % ARM
\cm &  % Cross
Page Table &
ASLR Bypass &
\xm &  % HW Mitigation
\cm \\ \hline  % SW Mitigation

Kemerlis et al. (2014) &
PTA &
Direct Map Exploitation &
\cm &  % Intel
\cm &  % AMD
\xm &  % ARM
\xm &  % Cross
Kernel Direct Mapping &
Privilege Escalation &
\xm &  % HW Mitigation
\cm \\ \hline  % SW Mitigation

Gras et al. (2018) &
PTA &
TLB Contention &
\cm &  % Intel
\xm &  % AMD
\xm &  % ARM
\xm &  % Cross
Translation Buffer &
Key Leakage &
\xm &  % HW Mitigation
\xm \\ \hline  % SW Mitigation

Hund et al. (2013) &
PTA, CTA &
Kernel ASLR Probing &
\cm &  % Intel
\cm &  % AMD
\xm &  % ARM
\cm &  % Cross
Page Tables \& Cache &
KASLR Bypass &
\xm &  % HW
\cm \\ \hline  % SW

Gras et al. (2017) &
PTA, CTA &
Evict+Time &
\cm &  % Intel
\cm &  % AMD
\xm &  % ARM
\cm &  % Cross
Page Table Cache &
ASLR Bypass &
\xm &  % HW
\xm \\ \hline  % SW

Gruss et al. (2019) &
PTA, ETA &
Page Cache Timing &
\cm &  % Intel
\cm &  % AMD
\cm &  % ARM
\cm &  % Cross
OS Page Cache &
ASLR Bypass &
\xm &  % HW
\cm \\ \hline  % SW

Schwarzl et al. (2023) &
ETA &
Decompression Timing &
\cm &  % Intel
\cm &  % AMD
\cm &  % ARM
\cm &  % Cross
Compressed Memory &
Data Leakage &
\xm &  % HW Mitigation
\xm \\ \hline  % SW Mitigation

Schwarz et al. (2017) &
ETA &
Timer Reconstruction &
\cm &  % Intel
\cm &  % AMD
\cm &  % ARM
\cm &  % Cross
JavaScript Timer APIs &
Timing Leakage &
\xm &  % HW Mitigation
\cm \\ \hline  % SW Mitigation

Van Goethem et al. (2020) &
ETA &
Timeless Timing &
\cm &  % Intel
\cm &  % AMD
\cm &  % ARM
\cm &  % Cross
Thread Scheduler &
Remote Secret Leak &
\xm &  % HW Mitigation
\cm \\ \hline  % SW Mitigation

Xiao \& Ainsworth (2023) &
ETA &
Parallel Timing Gadget &
\cm &  % Intel
\cm &  % AMD
\cm &  % ARM
\cm &  % Cross
Out-of-Order Execution &
Precision Timing &
\xm &  % HW Mitigation
\cm \\ \hline  % SW Mitigation

\end{tabular}
}
\end{table*}

\subsubsection{\textbf{Transient and Speculative Execution Attacks (TSEA)}}
Transient and speculative execution attacks exploit speculative execution, a CPU optimization technique where instructions are executed before their necessity is confirmed. Incorrect speculative paths leave traces in microarchitectural components, such as caches, which attackers analyze to infer sensitive data. The \textit{Spectre} and \textit{Meltdown}~\cite{kocher2020spectre, lipp2020meltdown} attacks were the first to demonstrate how speculative and out-of-order execution could be abused to leak sensitive information across isolation boundaries, exploiting microarchitectural side effects left behind by transient instructions. 

Building on these foundational vulnerabilities, \textit{Bhattacharyya et al.}~\cite{bhattacharyya2019smotherspectre} revealed how port contention channels could amplify speculative leakage across cores, whereas \textit{Islam et al.}~\cite{islam2019spoiler} demonstrated that speculative load hazards can enhance memory-based exploitation techniques such as Rowhammer and cache attacks. Subsequent research explored other prediction structures; for example,\textit{ Koruyeh et al.}~\cite{koruyeh2018spectre} leveraged RSB to extract sensitive information with high precision, and \textit{Van Bulck et al.}~\cite{van2018foreshadow} showed that transient execution could be used to compromise secure enclave boundaries in Intel SGX. More recently, \textit{Wikner et al.}~\cite{wikner2022retbleed} targeted return instructions on both x86-64 and ARM architectures, expanding the scope of speculation-based threats. \textit{Trujillo et al.}~\cite{trujillo2023inception} further advanced control-flow hijacking techniques by training microarchitectural buffers during transient execution. 

Recent studies have further demonstrated that transient execution vulnerabilities extend across a wide range of modern processors, including Intel, AMD, and ARM architectures. \textit{Shen et al.}~\cite{shen2024specrun} exploited speculative execution to remove constraints on transient instruction execution, exposing deeper control over speculative paths. Similarly, \textit{BranchSpectre}~\cite{chowdhuryy2021leaking} targeted branch prediction units to exfiltrate data through speculative updates. \textit{Ragab et al.}~\cite{ragab2024ghostrace} emphasized the widespread nature of these flaws by demonstrating speculative race conditions across all three major processor families. Focusing on ARM-based systems, \textit{Kim et al.}~\cite{kim2024tiktag} bypassed memory tagging extension (MTE) using speculative execution, revealing critical weaknesses in modern memory safety mechanisms. Meanwhile, \textit{Li et al.}~\cite{li2024indirector} leveraged high-precision Branch Target Injection (BTI) techniques to exploit speculative vulnerabilities specific to Intel CPUs.

\subsubsection{\textbf{Cache Timing Attacks (CTA)}}
Cache timing attacks exploit timing differences in accessing cache lines to reveal sensitive information. Foundational techniques such as \textit{Prime+Probe}~\cite{liu2015last} and \textit{Flush+Reload}~\cite{yarom2014flush+} monitor cache activity through precise timing measurements. \textit{Van Schaik et al.}~\cite{van2021cacheout} extended these techniques by leaking data through cache eviction patterns and \textit{Dall et al.}~\cite{dall2018cachequote} demonstrated how long-term secrets could be recovered from SGX environments using similar side channels. \textit{Canella et al.}~\cite{canella2019fallout} examined timing leaks in Meltdown-resistant CPUs caused by cache evictions, while \textit{Yarom et al.} identified side-channel vulnerabilities in constant-time cryptographic algorithms due to cache-bank conflicts~\cite{yarom2017cachebleed}. \textit{Gruss et al.}~\cite{gruss2015cache} introduced scalable methods to automate cache-based side-channel exploitation, and \textit{Liu et al.}~\cite{liu2015last} revealed critical vulnerabilities in multicore systems.

Adaptations of \textit{Prime+Probe} were shown to bypass browser-level defenses, even in constrained JavaScript environments~\cite{shusterman2021prime+}. \textit{Hund et al.}~\cite{hund2013practical} analyzed cache behavior to undermine kernel address space layout randomization (ASLR). \textit{Gruss et al.}~\cite{gruss2019page} demonstrated that file access patterns could be inferred through the page cache, and \textit{Disselkoen et al.}~\cite{disselkoen2017prime+} proposed a high-precision L3 cache attack using speculative aborts. Earlier studies by \textit{Oren et al.}~\cite{oren2015spy} and \textit{Aciicmez}~\cite{aciiccmez2007yet} extended cache timing attacks into shared and sandboxed environments, establishing their practicality in multi-user systems. Later research showed that these attacks could succeed without relying on page faults, by exploiting fine-grained cache access patterns in enclaved execution and shared memory contexts~\cite{van2017telling, yan2020cache}. Further, \textit{Zhang et al.}~\cite{zhang2012cross} revealed that shared caches in virtualized systems could be used to mount cross-VM side-channel attacks, exposing critical risks in cloud-based multi-tenant infrastructures.

\subsubsection{\textbf{Paging Timing Attacks (PTA)}}

Paging timing attacks exploit timing variations in memory management mechanisms, such as translation lookaside buffer behavior and page table updates, to infer sensitive information~\cite{gras2018translation, wang2017leaky}. These attacks exploit discrepancies in page access times, enabling attackers to extract data patterns from memory. These attacks, while timing-specific, are part of broader vulnerabilities in page table mechanisms that can be exploited in various ways, as explored under access pattern attacks. Additional research has shown that address translation behaviors can be exploited to bypass kernel-level protections such as Address Space Layout Randomization (ASLR)~\cite{gruss2016prefetch}, and that memory isolation mechanisms can be undermined through manipulation of rogue memory access patterns during kernel execution~\cite{kemerlis2014ret2dir}

\textit{Wang et al.}~\cite{wang2017leaky} demonstrated vulnerabilities in SGX, revealing how page table-based access patterns compromise secure enclaves. Similarly, \textit{Xu et al.}~\cite{xu2015controlled} utilized deterministic side channels by eliminating noise from page faults. Page table activity has been shown to leak architectural details of deep neural networks in shared environments~\cite{yan2020cache}. Enclave memory access patterns can also be exposed without page faults, enabling stealthy observation of execution behavior~\cite{van2017telling}. Additionally, TLB timing attacks have been used to bypass cache-based defenses, revealing systemic memory isolation flaws~\cite{gras2018translation}.

Further research expanded the scope of PTA vulnerabilities. A benchmark suite for evaluating cache vulnerability to timing attacks systematically assessed cache and paging vulnerabilities~\cite{deng2020benchmark}. Evaluation of cache attacks on ARM processors analyzed secure cache designs for ARM architectures, providing insights into mitigating timing attacks~\cite{deng2021evaluation}. The precise timed management of cache evictions has been proven to boost the reliability and precision of cache-based side-channel attacks, even against hardened cryptographic implementations~\cite{briongos2020cache}. Timing variations in page table caching have also revealed exploitable side channels in virtualized systems, highlighting risks in multi-tenant environments~\cite{zhang2014cache}.

\subsubsection{\textbf{Execution Timing Attacks (ETA)}}
Execution timing attacks exploit timing discrepancies during program execution to extract sensitive information. Unlike other timing-based attacks, ETA leverages broader timing variations caused by specific operations or architectural behaviors. Decompression timing has been shown to leak system-level data during memory compression operations~\cite{schwarzl2023practical}. Disturbance errors in DRAM, triggered by repeated row activations, enable attacks like \textit{Drammer}~\cite{van2016drammer} and \textit{Flipping Bits}~\cite{kim2014flipping} to break memory isolation. \textit{Rowhammer}~\cite{bhattacharya2016curious} induced faults have also been used to flip cryptographic exponent bits, compromising key integrity bits.

High-resolution timers enable microarchitectural side-channel attacks without elevated privileges, allowing fine-grained control over timing measurements~\cite{schwarz2017fantastic}. Timing-based information leakage can also occur remotely by leveraging concurrency in network protocols~\cite{van2020timeless}. \textit{Cheng et al.}~\cite{cheng2024evict+} presented that \textit{Evict+Spec+Time} exploits speculative execution to enhance cache-timing attacks, while timing the transient execution exposes vulnerabilities in Intel CPUs by utilizing transient execution timing~\cite{jin2023timing}. \textit{Xiao et al.}~\cite{xiao2023hacky} proposed a methodology entitled \textit{Hacky Racers} that uses instruction-level parallelism to create stealthy fine-grained timers. Moreover, coding practices introducing unintended timing differences remain a practical source of leaks~\cite{kholoosi2023empirical}.
\subsection{Access Pattern Attacks (APA)}
Access pattern attacks exploit predictable memory access behaviors to extract sensitive data or compromise system integrity~\cite{van2017telling}. These attacks leverage vulnerabilities in shared memory resources, such as page tables, secure enclaves, and memory deduplication mechanisms, to infer critical information~\cite{bosman2016dedup}. Unlike timing-based attacks, which rely on latency variations, APA focuses on observing memory access patterns to uncover operational traces that bypass traditional defenses. 

As cloud computing and virtualization become integral to modern infrastructure, APA presents a pressing threat to data confidentiality and system integrity~\cite{zhang2012cross}. These categories include secure enclave attacks (SEA), memory duplicate attacks (MDA), and page table entry attacks (PTEA), each targeting different layers of the system’s memory behavior to compromise confidentiality and isolation. In Table~\ref{tab:comparison-table-apa} we compare recent works that focused on APA-based SCAM.

SEA analyzes access patterns within trusted execution environments, such as Intel SGX and ARM TrustZone, to bypass isolation guarantees~\cite{wang2017leaky, xu2015controlled}. These attacks expose the limitations of trusted environments, where isolated memory can inadvertently leak sensitive operations. MDA exploits resource-sharing optimizations, such as merging identical memory pages, to infer private data or manipulate shared content~\cite{suzaki2011software}. PTEA reveals systemic flaws in memory management by manipulating page table structures to bypass isolation mechanisms and escalate privileges ~\cite{evtyushkin2016jump}.

% Please add the following required packages to your document preamble:
% \usepackage{multirow}
\begin{table*}[!ht]
\centering
\caption{Comparison of Different Access-Based Memory Side-Channel Attacks}
\label{tab:comparison-table-apa}
\resizebox{\textwidth}{!}{
\begin{tabular}{|c|c|c|c|c|c|c|c|c|c|c|}
\hline
\multirow{2}{*}{\textbf{Literature (Year)}} &
\multirow{2}{*}{\begin{tabular}[c]{@{}c@{}}\textbf{Attack}\\\textbf{Type}\end{tabular}} &
\multirow{2}{*}{\textbf{Methodology}} &
\multicolumn{4}{c|}{\textbf{Platform}} &
\multirow{2}{*}{\textbf{Target}} &
\multirow{2}{*}{\textbf{Impact}} &
\multicolumn{2}{c|}{\textbf{Mitigation}} \\ \cline{4-7} \cline{10-11}
& & & \textbf{Intel} & \textbf{AMD} & \textbf{ARM} & \textbf{Cross} & & & \textbf{HW} & \textbf{SW} \\ \hline\hline

Lee et al. (2017) &
SEA &
Branch Shadowing &
\cm &  % Intel
\xm &  % AMD
\xm &  % ARM
\xm &  % Cross
Branch Predictor &
Control Flow Leak &
\xm &  % HW Mitigation
\cm \\ \hline  % SW Mitigation

Murdock et al. (2020) &
SEA &
Voltage Fault Injection &
\cm &  % Intel
\xm &  % AMD
\xm &  % ARM
\xm &  % Cross
Voltage Interface &
Enclave Fault Injection &
\xm &  % HW Mitigation
\cm \\ \hline  % SW Mitigation

Gras et al. (2018) &
SEA &
TLB Profiling &
\cm &  % Intel
\xm &  % AMD
\xm &  % ARM
\xm &  % Cross
Translation Buffer &
EdDSA Key Recovery &
\xm &  % HW Mitigation
\cm \\ \hline  % SW Mitigation

Li et al. (2021) &
SEA &
Address Space Reuse &
\xm &  % Intel
\cm &  % AMD
\xm &  % ARM
\xm &  % Cross
TLB and Cache &
VM Memory Leak &
\xm &  % HW Mitigation
\cm \\ \hline  % SW Mitigation

Evtyushkin et al. (2016) &
SEA &
BTB Collision Probing &
\cm &  % Intel
\xm &  % AMD
\xm &  % ARM
\xm &  % Cross
Branch Target Buffer &
ASLR Bypass &
\cm &  % HW Mitigation
\cm \\ \hline  % SW Mitigation

Chen et al. (2019) &
SEA &
Branch Injection &
\cm &  % Intel
\xm &  % AMD
\xm &  % ARM
\xm &  % Cross
Branch Predictor &
SGX Key Leakage &
\xm &  % HW Mitigation
\cm \\ \hline  % SW Mitigation

Schwarz et al. (2019) &
SEA &
Fill Buffer Sampling &
\cm &  % Intel
\xm &  % AMD
\xm &  % ARM
\xm &  % Cross
Fill Buffer &
SGX Data Leakage &
\xm &  % HW Mitigation
\cm \\ \hline  % SW Mitigation

Evtyushkin et al. (2018) &
SEA &
Directional Probing &
\cm &  % Intel
\xm &  % AMD
\xm &  % ARM
\xm &  % Cross
Branch Predictor &
Control Flow Leak &
\xm &  % HW Mitigation
\cm \\ \hline  % SW Mitigation

Van Bulck et al. (2017) &
PTEA &
PTE Monitoring &
\cm &  % Intel
\xm &  % AMD
\xm &  % ARM
\xm &  % Cross
Page Table Entry &
Access Pattern Leakage &
\xm &  % HW Mitigation
\cm \\ \hline  % SW Mitigation

Van Schaik et al. (2017) &
PTEA &
Page Table Cache Probing &
\cm &  % Intel
\cm &  % AMD
\cm &  % ARM
\cm &  % Cross
Page Walk Caches &
Microarchitecture mapping &
\xm &  % HW Mitigation
\xm \\ \hline  % SW Mitigation

Notselwyn (2024) &
PTEA &
Page Directory Confusion &
\cm &  % Intel
\cm &  % AMD
\xm &  % ARM
\xm &  % Cross
Page Tables Entry &
Kernel Memory Access &
\xm &  % HW Mitigation
\cm \\ \hline  % SW Mitigation

Yoshioka et al. (2024) &
PTEA &
Global Bit Flipping &
\cm &  % Intel
\xm &  % AMD
\xm &  % ARM
\xm &  % Cross
Page Table Entry &
Memory Access Violation &
\xm &  % HW Mitigation
\cm \\ \hline  % SW Mitigation

Moghimi et al. (2020) &
PTEA &
Instruction Fault Tracing &
\cm &  % Intel
\xm &  % AMD
\xm &  % ARM
\xm &  % Cross
Page Table Entry &
Instruction Control Flow &
\xm &  % HW Mitigation
\cm \\ \hline  % SW Mitigation

Bosman et al. (2016) &
MDA &
Deduplication Probing &
\cm &  % Intel
\cm &  % AMD
\xm &  % ARM
\cm &  % Cross
Shared Pages &
ASLR Bypass &
\xm &  % HW Mitigation
\cm \\ \hline  % SW Mitigation

Zhang et al. (2012) &
MDA &
Prime+Probe &
\cm &  % Intel
\cm &  % AMD
\xm &  % ARM
\cm &  % Cross
L3 Cache &
ElGamal Key Extraction &
\xm &  % HW Mitigation
\cm \\ \hline  % SW Mitigation

Suzaki et al. (2011) &
MDA &
Deduplication Probing &
\cm &  % Intel
\cm &  % AMD
\xm &  % ARM
\cm &  % Cross
Shared Pages &
App/VM Discovery &
\xm &  % HW Mitigation
\cm \\ \hline  % SW Mitigation

Lindemann et al. (2018) &
MDA &
Deduplication Timing &
\cm &  % Intel
\cm &  % AMD
\xm &  % ARM
\xm &  % Cross
Shared Pages &
Application Detection &
\xm &  % HW Mitigation
\xm \\ \hline  % SW Mitigation

Schwarzl et al. (2021) &
MDA &
Remote Deduplication &
\cm &  % Intel
\cm &  % AMD
\xm &  % ARM
\cm &  % Cross
Shared Pages &
Remote Data Disclosure &
\xm &  % HW Mitigation
\cm \\ \hline  % SW Mitigation

Barresi et al. (2015) &
MDA &
Deduplication Probing &
\cm &  % Intel
\cm &  % AMD
\xm &  % ARM
\cm &  % Cross
Shared Pages &
ASLR Bypass &
\xm &  % HW Mitigation
\cm \\ \hline  % SW Mitigation

Kim et al. (2021) &
MDA &
Deduplication Timing &
\cm &  % Intel
\cm &  % AMD
\xm &  % ARM
\cm &  % Cross
Shared Pages &
KASLR Bypass &
\xm &  % HW Mitigation
\cm \\ \hline  % SW Mitigation

Suzaki et al. (2011) &
MDA &
Deduplication Timing &
\cm &  % Intel
\cm &  % AMD
\xm &  % ARM
\cm &  % Cross
Shared Pages &
App Presence Detection &
\xm &  % HW Mitigation
\cm \\ \hline  % SW Mitigation

Suzaki et al. (2013) &
MDA &
Deduplication Timing &
\cm &  % Intel
\cm &  % AMD
\xm &  % ARM
\cm &  % Cross
Shared Pages &
App Presence Detection &
\xm &  % HW Mitigation
\cm \\ \hline  % SW Mitigation

\end{tabular}
}
\end{table*}

\subsubsection{\textbf{Secure Enclave Attacks (SEA)}}
Trusted execution environments (TEE), such as Intel’s Software Guard Extensions (SGX), aim to isolate sensitive computations from untrusted system components, but have been shown vulnerable to a variety of microarchitectural attacks~\cite{zheng2021survey}. As a subcategory of access pattern analysis attacks, SEA exploits memory access behaviors to compromise isolation boundaries, targeting weaknesses in branch prediction, speculative execution, operating system interaction and fault handling mechanisms~\cite{van2017telling, lee2017inferring, murdock2020plundervolt}. By observing and manipulating access patterns, attackers can undermine the integrity of these trusted environments, exposing critical data even in systems designed for high-assurance security.

One prominent class of SEA focuses on branch prediction, where speculative control flows are manipulated to leak information from within the enclave via microarchitectural side channels~\cite{lee2017inferring, evtyushkin2018branchscope}. In parallel, operating system-level attacks exploit the enclave’s reliance on untrusted OS services, using deterministic side channels and timing variations to subvert SGX’s isolation guarantees~\cite{xu2015controlled, kemerlis2014ret2dir, gras2018tlbleed}. Hardware-based vectors, such as shared caches and transient execution faults, have also been shown to compromise enclave and kernel boundaries, revealing data through fine-grained access pattern observation~\cite{yan2020cache}. Building on this, transient execution techniques demonstrate how speculative CPU behaviors can be exploited to bypass architectural protections and expose enclave-resident secrets~\cite{van2018foreshadow, koruyeh2018spectre, van2019ridl, schwarz2019zombieload}. Fault-based methods similarly introduce hardware-level errors or corrupt execution states to extract sensitive data from within secure enclaves~\cite{murdock2020plundervolt, cui2021smashex, sridhara2024sigy, cloosters2020teerex}.

\subsubsection{\textbf{Page Table Entry Attacks (PTEA)}}
Page table entry attacks exploit flaws in the structure and management of page tables, a core component of modern memory systems~\cite{gras2018translation, van2017revanc}. By manipulating these tables, attackers can compromise memory isolation, gain unauthorized access, and escalate privileges~\cite{xu2015controlled}. As computing increasingly relies on virtualized and shared environments, PTEA highlights critical architectural vulnerabilities.

Page tables map virtual memory to physical memory and are responsible for enforcing separation between user and kernel spaces. However, these mappings can be abused to bypass memory protections. For instance, prior work has shown that exploiting the separation between user and kernel address spaces can undermine kernel address space layout randomization (KASLR), exposing sensitive memory regions~\cite{evtyushkin2016jump}. Additionally, manipulation of user-space page tables has been used to gain unauthorized kernel memory access by targeting slab allocation strategies~\cite{van2017telling}. PTEA techniques have also enabled privilege escalation by corrupting kernel mappings~\cite{flippingpages2024} and modifying page table entries to create writable kernel regions~\cite{wicherski2013}. Beyond direct memory control, attacks targeting metadata, such as the global bit, have been shown to facilitate malicious page sharing and arbitrary code execution~\cite{yoshioka2024gbhammer}.

These techniques underscore how low-level memory control and metadata manipulation can bypass traditional protections—threats that now extend beyond classical system architectures. In virtualized environments, decrypted page tables have enabled code injection under AMD SEV~\cite{li2021crossline}, while in secure enclaves, attackers have used access patterns to infer instruction-level control flow~\cite{moghimi2020copycat}. Furthermore, dynamic manipulation of protection attributes has revealed runtime vulnerabilities that undermine even advanced memory isolation mechanisms~\cite{parida2021pagedumper}.

\subsubsection{\textbf{Memory Deduplication Attacks (MDA)}}
Memory deduplication attacks exploit memory optimization techniques commonly used in modern cloud and virtualization platforms, where identical memory pages across virtual machines are merged to enhance efficiency~\cite{bosman2016dedup}. While this strategy improves performance, it introduces exploitable side channels that compromise system security~\cite{suzaki2011software}. As a subcategory of Access Pattern Attacks, MDA leverages shared memory behavior to infer sensitive information, undermining the isolation guarantees fundamental to virtualized environments~\cite{zhang2012cross}.

Notably, memory deduplication has been used to bypass Kernel Address Space Layout Randomization (ASLR) by crafting controlled memory patterns and monitoring deduplication activity to infer co-resident virtual machine layouts~\cite{barresi2015cain}. Building on this principle, side-channel methods have revealed the presence of specific applications in neighboring VMs, raising serious concerns in multi-tenant architectures~\cite{lindemann2018memory}. These vulnerabilities extend beyond local threats: researchers have shown that remote attackers can infer memory content by analyzing timing variations in HTTP responses, highlighting the broad attack surface of deduplication-based leakage~\cite{schwarzl2021remote}.

Even hardened systems, such as kernel page table isolation (KPTI) enabled Linux VMs, have proven susceptible, with deduplication-based channels successfully breaking ASLR protections~\cite{kim2021breaking}. The impact on cryptographic libraries is equally concerning; shared memory optimizations have been exploited to extract sensitive keys and other secure data~\cite{suzaki2013implementation}. Advanced exploitation techniques demonstrate the adaptability of MDA, revealing cross-VM leakage and even enabling attacks from within sandboxed environments like JavaScript~\cite{bosman2016dedup}.
\subsection{Signal-Based Attacks (SBA)}

Signal-based attacks exploit physical emanations, such as power consumption, thermal variations, or electromagnetic emissions, to extract sensitive information from computing systems~\cite{mangard2008power}. With advances in hardware technologies and growing reliance on interconnected systems, these vulnerabilities are increasingly evident in critical domains, posing significant risks to modern infrastructures such as IoT devices, cloud platforms, and cryptographic hardware~\cite{ignatius2024power}. For example, power analysis attacks have been used to compromise IoT devices in smart homes~\cite{dione2023hardware}, while electromagnetic leakage has exposed cryptographic keys during financial transactions~\cite{frieslaar2018developing}. 

Signal-based attacks underscore the unintended security consequences of hardware performance optimizations.
SBA encompasses three primary subcategories based on the physical phenomena they exploit. These include Power Analysis Attacks (PAA), Thermal Attacks (TA), and Electromagnetic Leakage Attacks (ELA), each leveraging distinct physical side channels to extract sensitive system information. PAA uses power consumption patterns to infer cryptographic keys or system states, with techniques such as simple power analysis (SPA) and differential power analysis (DPA) being foundational~\cite{mangard2008power, kocher1999differential}. TA exploits heat dissipation to breach security boundaries, leveraging temperature variations in microcontrollers or cryptographic devices~\cite{hutter2014temperature}. ELA analyzes electromagnetic emissions from system operations, revealing internal hardware states~\cite{agrawal2003multi}. Table~\ref{tab:comparison-table-sba} compares SBA-based SCAM techniques as identified in the current literature.

% Please add the following required packages to your document preamble:
% \usepackage{multirow}
\begin{table*}[!ht]
\centering
\caption{Comparison of Different Signal-Based Memory Side-Channel Attacks}
\label{tab:comparison-table-sba}
\resizebox{\textwidth}{!}{
\begin{tabular}{|c|c|c|c|c|c|c|c|c|c|c|}
\hline
\multirow{2}{*}{\textbf{Literature (Year)}} &
\multirow{2}{*}{\begin{tabular}[c]{@{}c@{}}\textbf{Attack}\\\textbf{Type}\end{tabular}} &
\multirow{2}{*}{\textbf{Methodology}} &
\multicolumn{4}{c|}{\textbf{Platform}} &
\multirow{2}{*}{\textbf{Target}} &
\multirow{2}{*}{\textbf{Impact}} &
\multicolumn{2}{c|}{\textbf{Mitigation}} \\ \cline{4-7} \cline{10-11}
& & & \textbf{Intel} & \textbf{AMD} & \textbf{ARM} & \textbf{Cross} & & & \textbf{HW} & \textbf{SW} \\ \hline\hline

Mangard et al. (2008) &
PAA &
Power Trace Analysis &
\xm &  % Intel
\xm &  % AMD
\xm &  % ARM
\cm &  % Cross
Smart Card Logic &
Secret Key Leak &
\xm &  % HW Mitigation
\cm \\ \hline  % SW Mitigation

Ignatius et al. (2024) &
PAA &
Power Trace Recovery &
\xm &  % Intel
\xm &  % AMD
\xm &  % ARM
\cm &  % Cross
RISC-V AES Core &
Key Leak &
\xm &  % HW Mitigation
\cm \\ \hline  % SW Mitigation

Brier et al. (2004) &
PAA &
Correlation Power Analysis &
\xm &  % Intel
\xm &  % AMD
\xm &  % ARM
\cm &  % Cross
Cryptographic Chips &
Key Leak &
\xm &  % HW Mitigation
\cm \\ \hline  % SW Mitigation

Srivastava et al. (2024) &
PAA &
RTL Power Analysis &
\xm &  % Intel
\xm &  % AMD
\xm &  % ARM
\cm &  % Cross
Crypto RTL Design &
Module Leak &
\cm &  % HW Mitigation
\cm \\ \hline  % SW Mitigation

Chawla et al. (2023) &
PAA &
Software Power Telemetry &
\xm &  % Intel
\xm &  % AMD
\xm &  % ARM
\cm &  % Cross
Apple M1/M2 SoC &
Key Leak &
\xm &  % HW Mitigation
\cm \\ \hline  % SW Mitigation

Bronchain et al. (2021) &
PAA &
Single Trace Correlation &
\xm &  % Intel
\xm &  % AMD
\xm &  % ARM
\cm &  % Cross
ATMega AES Core &
Key Leak &
\xm &  % HW Mitigation
\cm \\ \hline  % SW Mitigation

Das et al. (2019) &
PAA &
Cross Device Profiling &
\xm &  % Intel
\xm &  % AMD
\xm &  % ARM
\cm &  % Cross
AVR AES Core &
Key Leak &
\xm &  % HW Mitigation
\cm \\ \hline  % SW Mitigation

O’Flynn \& d’Eon (2018) &
PAA &
Power and Glitching &
\xm &  % Intel
\xm &  % AMD
\xm &  % ARM
\cm &  % Cross
CAN Microcontroller &
Key and Firmware &
\xm &  % HW Mitigation
\cm \\ \hline  % SW Mitigation

Kocher et al. (1999) &
PAA &
Differential Power Analysis &
\xm &  % Intel
\xm &  % AMD
\xm &  % ARM
\cm &  % Cross
Cryptographic Chips &
Key Leak &
\xm &  % HW Mitigation
\cm \\ \hline  % SW Mitigation

Wang et al. (2022) &
PAA &
DVFS Timing Exploitation &
\cm &  % Intel
\cm &  % AMD
\xm &  % ARM
\cm &  % Cross
x86 DVFS Mechanism &
Key Leak &
\xm &  % HW Mitigation
\cm \\ \hline  % SW Mitigation

Kim \& Shin (2022) &
TA &
Thermal Sensor Profiling &
\cm &  % Intel
\xm &  % AMD
\xm &  % ARM
\xm &  % Cross
CPU Thermal Sensor &
KASLR Bypass &
\xm &  % HW Mitigation
\cm \\ \hline  % SW Mitigation

Zhang et al. (2024) &
TA &
Thermal Interrupt Profiling &
\cm &  % Intel
\xm &  % AMD
\xm &  % ARM
\xm &  % Cross
Intel CPU Interrupts &
KASLR Bypass &
\xm &  % HW Mitigation
\cm \\ \hline  % SW Mitigation

Aljuffri et al. (2021) &
TA &
Thermal Signal Profiling &
\xm &  % Intel
\xm &  % AMD
\xm &  % ARM
\cm &  % Cross
RSA Microcontroller &
Key Leak &
\xm &  % HW Mitigation
\cm \\ \hline  % SW Mitigation

Hutter \& Schmidt (2014) &
TA &
Temperature Profiling &
\xm &  % Intel
\xm &  % AMD
\xm &  % ARM
\cm &  % Cross
CMOS Smart Card &
RSA Key Leak &
\xm &  % HW Mitigation
\cm \\ \hline  % SW Mitigation

Agrawal et al. (2003) &
ELA &
Multi Channel SCA &
\xm &  % Intel
\xm &  % AMD
\xm &  % ARM
\cm &  % Cross
Smart Card SoC &
Key Leak &
\xm &  % HW Mitigation
\cm \\ \hline  % SW Mitigation

Genkin et al. (2016) &
ELA &
EM Signal Profiling &
\cm &  % Intel
\xm &  % AMD
\xm &  % ARM
\xm &  % Cross
ECDH Module &
Key Leak &
\xm &  % HW Mitigation
\cm \\ \hline  % SW Mitigation

Genkin et al. (2014) &
ELA &
Acoustic Signal Profiling &
\cm &  % Intel
\xm &  % AMD
\xm &  % ARM
\xm &  % Cross
RSA Module &
Key Leak &
\xm &  % HW Mitigation
\cm \\ \hline  % SW Mitigation

Heyszl et al. (2012) &
ELA &
Localized EM Probing &
\xm &  % Intel
\xm &  % AMD
\xm &  % ARM
\cm &  % Cross
FPGA ECC Module &
Key Leak &
\xm &  % HW Mitigation
\cm \\ \hline  % SW Mitigation

Frieslaar \& Irwin (2018) &
ELA &
EM Profiling &
\xm &  % Intel
\xm &  % AMD
\cm &  % ARM
\cm &  % Cross
Pi AES Core &
Key Extraction &
\xm &  % HW Mitigation
\cm \\ \hline  % SW Mitigation

Batina et al. (2019) &
ELA &
EM Profiling &
\xm &  % Intel
\xm &  % AMD
\xm &  % ARM
\cm &  % Cross
Neural Net MCU &
Architecture Leak &
\xm &  % HW Mitigation
\cm \\ \hline  % SW Mitigation

Sayakkara et al. (2019) &
ELA &
EM Event Profiling &
\xm &  % Intel
\xm &  % AMD
\xm &  % ARM
\cm &  % Cross
IoT Microcontroller &
Trace Leak &
\xm &  % HW Mitigation
\xm \\ \hline  % SW Mitigation

Danial et al. (2021) &
ELA &
Cross-Device EM Profiling &
\xm &  % Intel
\xm &  % AMD
\xm &  % ARM
\cm &  % Cross
Embedded AES Core &
Key Leak &
\xm &  % HW Mitigation
\cm \\ \hline  % SW Mitigation

Yu et al. (2020) &
ELA &
EM Model Extraction &
\xm &  % Intel
\xm &  % AMD
\xm &  % ARM
\cm &  % Cross
FPGA DNN Accelerator &
Model Leak &
\xm &  % HW Mitigation
\cm \\ \hline  % SW Mitigation

\end{tabular}
}
\end{table*}

\subsubsection{\textbf{Power Analysis Attacks (PAA)}}
Power analysis attacks exploit variations in power consumption to extract sensitive information such as cryptographic keys, threatening the confidentiality of secure computations~\cite{mangard2008power}. As computing systems become increasingly dependent on hardware optimization, PAA exposes critical vulnerabilities across cryptographic implementations. By analyzing power traces, attackers uncover patterns linked to internal computations. Differential Power Analysis (DPA) by \textit{Kocher et al.}~\cite{kocher1999differential} demonstrated how subtle power trace differences reveal encryption keys, challenging cryptographic security. Simple Power Analysis (SPA) observes power consumption to infer cryptographic operations, exposing key-dependent variations in AES or modular exponentiation~\cite{mangard2008power}. Correlation Power Analysis (CPA) refines this approach by statistically linking measured power traces to hypothetical key models~\cite{brier2004correlation}. 

Modern applications of PAA highlight its versatility and growing relevance across sectors. For example, power analysis has been used to compromise IoT devices in healthcare and smart home environments~\cite{liao2020security}, while early-stage RTL vulnerability detection is now possible using tools like \textit{SCAR}~\cite{srivastava2024scar}. In the consumer hardware domain, telemetry-based techniques have extracted cryptographic keys from Apple M1/M2 processors~\cite{chawla2023power}. Similarly, attacks on neural network accelerators have been shown to leak training data and model parameters, posing serious threats to AI confidentiality~\cite{batina2019csi}. Recent advances in single-trace power analysis enable attackers to recover secrets using just one measurement, significantly increasing stealth and efficiency~\cite{bronchain2021give}. Furthermore, targeted attacks on open architectures such as RISC-V~\cite{das2019x} and embedded automotive controllers~\cite{oflynn2018can} demonstrate the continued feasibility of PAA across both customizable and resource-constrained platforms.

\subsubsection{\textbf{Thermal Attacks (TA)}}
Thermal attacks exploit heat dissipation patterns in computing devices to infer sensitive information or compromise system security. As systems become more compact and energy efficient, thermal emissions have emerged as a growing vector for side-channel exploitation~\cite{hutter2014temperature}. Temperature variations caused by cryptographic operations or computational workloads can create measurable side channels, enabling attackers to extract data such as encryption keys~\cite{hutter2014temperature}. Building on these foundations, recent research has demonstrated practical applications of thermal side channels across different environments. Thermal variations across CPU cores have been used to bypass address space layout randomization (ASLR) protections by analyzing heat distribution patterns~\cite{kim2022thermalbleed}, and interrupt-driven temperature manipulation enables advanced timing-based exploits~\cite{zhang2024thermalscope}. These techniques are particularly threatening in constrained or shared environments such as IoT devices and cloud-hosted platforms, where limited thermal isolation leaves systems more vulnerable~\cite{aljuffri2021applying, kim2022thermalbleed}.

More advanced approaches have leveraged thermal residue to infer user inputs like typed passwords~\cite{bekaert2022thermal}, and have exploited temperature-dependent behavior in GPUs to leak sensitive computations~\cite{liu2019side}. Emerging hardware designs such as 3D integrated circuits (3D ICs) have also been shown to be susceptible to thermal-based key leakage~\cite{benelhaouare2024mitigating}. Additionally, remote monitoring techniques, using heat-sensitive cameras or probes, further expand the reach of these attacks beyond physical contact~\cite{zhang2024thermalscope}.

\subsubsection{\textbf{Electromagnetic Leakage Attacks (ELA)}}

Electromagnetic leakage attacks exploit unintentional electromagnetic (EM) emissions from computing devices to extract sensitive information~\cite{agrawal2003side, genkin2014rsa}. These emissions, generated by hardware operations, create a side channel that attackers analyze to compromise cryptographic systems or infer confidential data, posing significant risks in sectors such as finance, healthcare, and automotive. \textit{Agrawal et al.}~\cite{agrawal2003side} introduced foundational methodologies for assessing EM vulnerabilities, establishing a basis for understanding subsequent advancements. Practical demonstrations, such as \textit{Genkin et al.}'s~\cite{genkin2016ecdh} extraction of ECDH keys using low-bandwidth EM attacks, highlighted the feasibility of real-world cryptographic leakage. Localized EM analysis further emphasized the precision of such attacks, particularly in constrained cryptographic implementations~\cite{heyszl2012localized}.

Building on these early findings, modern advances have significantly expanded the scope and sophistication of EM-based threats. For instance, \textit{Wang et al.}~\cite{wang2022hertzbleed} transformed a power-based vulnerability into a remote timing attack using EM leakage, showcasing the adaptability of EM exploitation techniques. IoT systems operating in multi-user environments face increasing exposure, with EM side channels revealing cryptographic weaknesses in constrained platforms~\cite{sayakkara2019leveraging}. Similarly, cross-device EM leakage studies have uncovered vulnerabilities in shared hardware, emphasizing risks in multi-tenant and virtualized environments~\cite{danial2021x}. High-resolution EM analysis has been used to pinpoint weaknesses in secure encryption schemes, demonstrating the power of precise EM measurements in exposing sensitive operations~\cite{das2020electromagnetic}. In practice, real-time EM leakage attacks have been implemented in automotive systems, where unintended emissions are leveraged to infer sensitive control or diagnostic data~\cite{kinugawa2017evaluation}. Emerging hardware accelerators used in AI workloads have also become targets, with attackers leveraging EM side channels to infer internal neural network operations and parameters~\cite{yu2020deepem}.
\subsection{Fault Injection Attacks (FIA)}
Fault Injection Attacks exploit disruptions in hardware or software to uncover vulnerabilities, compromise system integrity, and extract sensitive information. These attacks target computing systems' reliability, exposing weaknesses in devices from personal computers to IoT systems. As hardware optimizations and interconnected systems proliferate, the relevance of FIA continues to grow. FIA is categorized into three subtypes: DRAM Vulnerability Attacks (DVA), which induce memory corruption through hardware-level faults~\cite{mutlu2019rowhammer}; Cold Boot Attacks (CBA), which recover residual data from volatile memory after power loss~\cite{halderman2009lest}; and Peripheral DMA Attacks (PDA), leveraging DMA vulnerabilities for unauthorized access~\cite{markettos2019thunderclap}. Table~\ref{tab:comparison-table-fault} presents a comprehensive comparison of current literature that discusses fault injection attacks.

Over time, fault injection techniques have evolved from simple bit-flipping to advanced methods that bypass hardware defenses. DVA-based methods like row-based hammering exploit charge disturbance effects to induce memory bit-flips~\cite{mutlu2019rowhammer}, while newer approaches bypass mitigation techniques~\cite{frigo2020trrespass}. CBA has been used to extract encryption keys and sensitive data after power cycles~\cite{halderman2009lest}, and extended to recover machine learning models from volatile memory. PDA exploits DMA interfaces to manipulate system memory, posing a major threat in modern systems with PCIe-based peripherals~\cite{markettos2019thunderclap, de2023forensic}.

% Please add the following required packages to your document preamble:
% \usepackage{multirow}
\begin{table*}[!ht]
\centering
\caption{Comparison of Fault Injection-Based Memory Side-Channel Attacks}
\label{tab:comparison-table-fault}
\resizebox{\textwidth}{!}{
\begin{tabular}{|c|c|c|c|c|c|c|c|c|c|c|}
\hline
\multirow{2}{*}{\textbf{Literature (Year)}} &
\multirow{2}{*}{\begin{tabular}[c]{@{}c@{}}\textbf{Attack}\\\textbf{Type}\end{tabular}} &
\multirow{2}{*}{\textbf{Methodology}} &
\multicolumn{4}{c|}{\textbf{Platform}} &
\multirow{2}{*}{\textbf{Target}} &
\multirow{2}{*}{\textbf{Impact}} &
\multicolumn{2}{c|}{\textbf{Mitigation}} \\ \cline{4-7} \cline{10-11}
& & & \textbf{Intel} & \textbf{AMD} & \textbf{ARM} & \textbf{Cross} & & & \textbf{HW} & \textbf{SW} \\ \hline\hline

Mutlu et al. (2019) &
DVA &
DRAM Row Disturbance &
\cm &  % Intel
\cm &  % AMD
\cm &  % ARM
\cm &  % Cross
DRAM Memory Cell &
Privilege Escalation &
\cm &  % HW Mitigation
\cm \\ \hline  % SW Mitigation

Jattke et al. (2024) &
DVA &
Precision Row Cycling &
\xm &  % Intel
\cm &  % AMD
\xm &  % ARM
\cm &  % Cross
DDR4/DDR5 on AMD Zen &
Bit Flips \& Escalation &
\cm &  % HW Mitigation
\xm \\ \hline  % SW Mitigation

Gruss et al. (2016) &
DVA &
Eviction based Hammering &
\cm &  % Intel
\cm &  % AMD
\xm &  % ARM
\cm &  % Cross
DRAM in Browser Context &
Privilege Escalation &
\xm &  % HW Mitigation
\cm \\ \hline  % SW Mitigation

Tatar et al. (2018) &
DVA &
Remote Rowhammer Trigger &
\cm &  % Intel
\cm &  % AMD
\xm &  % ARM
\cm &  % Cross
DRAM via RDMA &
Remote Execution &
\xm &  % HW Mitigation
\cm \\ \hline  % SW Mitigation

Jattke et al. (2022) &
DVA &
Pattern based Hammering &
\cm &  % Intel
\cm &  % AMD
\xm &  % ARM
\cm &  % Cross
DDR4 DRAM Modules &
Bit Flips Escalation &
\cm &  % HW Mitigation
\xm \\ \hline  % SW Mitigation

Van Der Veen et al. (2016) &
DVA &
Predictive Memory Hammering &
\xm &  % Intel
\xm &  % AMD
\cm &  % ARM
\cm &  % Cross
Mobile DRAM on ARM &
Privilege Escalation &
\xm &  % HW Mitigation
\cm \\ \hline  % SW Mitigation

Islam et al. (2019) &
DVA &
Speculative Address Profiling &
\cm &  % Intel
\xm &  % AMD
\xm &  % ARM
\xm &  % Cross
Load Buffer Logic &
Memory Bit Flip &
\xm &  % HW Mitigation
\cm \\ \hline  % SW Mitigation

Bhattacharya et al. (2016) &
DVA &
Controlled Bit Flip Induction &
\cm &  % Intel
\xm &  % AMD
\xm &  % ARM
\xm &  % Cross
DRAM RSA Key Region &
RSA Key Leak &
\xm &  % HW Mitigation
\cm \\ \hline  % SW Mitigation

Kwong et al. (2020) &
DVA &
Data Dependent Rowhammer &
\cm &  % Intel
\cm &  % AMD
\xm &  % ARM
\cm &  % Cross
DRAM Adjacent Rows &
RSA Key Leak &
\xm &  % HW Mitigation
\cm \\ \hline  % SW Mitigation

Frigo et al. (2020) &
DVA &
Multi-Row Hammering &
\cm &  % Intel
\cm &  % AMD
\xm &  % ARM
\cm &  % Cross
DDR4 DRAM Modules &
Bit Flips \& Escalation &
\cm &  % HW Mitigation
\xm \\ \hline  % SW Mitigation

Markettos et al. (2019) &
PDA &
DMA Memory Injection &
\cm &  % Intel
\cm &  % AMD
\xm &  % ARM
\cm &  % Cross
IOMMU \& System Memory &
Privilege Escalation &
\xm &  % HW Mitigation
\cm \\ \hline  % SW Mitigation

Frisk et al. (2016) &
PDA &
PCIe DMA Injection &
\cm &  % Intel
\cm &  % AMD
\xm &  % ARM
\cm &  % Cross
Kernel Memory (RAM) &
Privilege Escalation &
\xm &  % HW Mitigation
\cm \\ \hline  % SW Mitigation

Assumpção et al. (2023) &
PDA &
Debug Link Injection &
\cm &  % Intel
\xm &  % AMD
\xm &  % ARM
\xm &  % Cross
Intel DCI Memory Path &
Key Leak &
\xm &  % HW Mitigation
\xm \\ \hline  % SW Mitigation

Frisk et al. (2021) &
PDA &
PCIe DMA Injection &
\cm &  % Intel
\cm &  % AMD
\xm &  % ARM
\cm &  % Cross
System RAM (via PCIe) &
Memory Dump &
\xm &  % HW Mitigation
\cm \\ \hline  % SW Mitigation

Halderman et al. (2009) &
CBA &
RAM Content Dumping &
\cm &  % Intel
\cm &  % AMD
\xm &  % ARM
\cm &  % Cross
RAM Encryption Storage &
Key Leak &
\xm &  % HW Mitigation
\cm \\ \hline  % SW Mitigation

Won et al. (2021) &
CBA &
Cold Boot Dumping &
\cm &  % Intel
\xm &  % AMD
\xm &  % ARM
\cm &  % Cross
EdgeML Accelerator RAM &
Model Parameter Leak &
\xm &  % HW Mitigation
\cm \\ \hline  % SW Mitigation

Villanueva et al. (2020) &
CBA &
Partial Key Recovery &
\cm &  % Intel
\cm &  % AMD
\xm &  % ARM
\cm &  % Cross
RAM (LUOV Private Key) &
Private Key Leak &
\xm &  % HW Mitigation
\cm \\ \hline  % SW Mitigation

Wetzels et al. (2014) &
CBA &
Cold Boot Dumping &
\cm &  % Intel
\cm &  % AMD
\xm &  % ARM
\cm &  % Cross
Volatile RAM &
Cipher Key Leak &
\xm &  % HW Mitigation
\cm \\ \hline  % SW Mitigation

Zimerman et al. (2021) &
CBA &
Neural Key Recovery &
\cm &  % Intel
\cm &  % AMD
\xm &  % ARM
\cm &  % Cross
DRAM (AES Key Region) &
AES Key Leak &
\xm &  % HW Mitigation
\cm \\ \hline  % SW Mitigation

Banegas et al. (2023) &
CBA &
Hybrid Key Enumeration &
\cm &  % Intel
\cm &  % AMD
\xm &  % ARM
\cm &  % Cross
DRAM Key Material &
Secret Key Leak &
\xm &  % HW Mitigation
\cm \\ \hline  % SW Mitigation

\end{tabular}
}
\end{table*}

\subsubsection{\textbf{DRAM Vulnerability Attacks (DVA)}} DRAM vulnerability attacks exploit weaknesses in dynamic random-access memory to induce memory corruption, gain unauthorized access, or extract sensitive information. These attacks exploit the physical characteristics of DRAM cells. Despite its optimization for high-density storage, DRAM remains vulnerable to disturbance errors~\cite{mutlu2019rowhammer}. Initial research showed that repeatedly accessing a single memory row could induce bit flips in adjacent rows, undermining data integrity~\cite{mutlu2019rowhammer}. Later developments bypassed built-in mitigation techniques such as Target Row Refresh (TRR), exposing persistent hardware limitations~\cite{frigo2020trrespass}. Building on these foundations, researchers introduced deterministic bit-flip patterns for greater reliability~\cite{jattke2022blacksmith, jattke2024zenhammer}, and demonstrated successful exploitation even in constrained environments like JavaScript engines and remote systems~\cite{gruss2016rowhammer, tatar2018throwhammer}.

Beyond traditional platforms, mobile devices have also proven vulnerable. \textit{Rowhammer}-style attacks have been executed on ARM-based phones~\cite{van2016drammer}, and power-based analysis techniques have been used to detect execution patterns in DRAM manipulation~\cite{cohen2022hammerscope}. Recent work has expanded these attacks into new threat models: for example, the faults induced by \textit{Rowhammer} have been shown to impact cryptographic computations~\cite{bhattacharya2016curious}, and speculative behavior has been linked to the amplification of DRAM vulnerabilities~\cite{islam2019spoiler, bhattacharyya2019smotherspectre}. Fault injection has even been transformed into a data exfiltration channel, using access pattern leakage in shared memory environments~\cite{kwong2020rambleed}.

\subsubsection{\textbf{Cold Boot Attacks (CBA)}}
Cold boot attacks exploit the residual data retention properties of Dynamic Random-Access Memory at low temperatures to recover sensitive information, such as cryptographic keys, even after the system is powered off. These attacks exploit the fact that data in DRAM does not disappear immediately after power-off and can be preserved longer using freezing techniques. \textit{Halderman et al}~\cite{halderman2009lest} first demonstrated how data remanence in DRAM could be exploited to recover encryption keys and other sensitive information. Subsequent studies built on this foundation to target more advanced applications, including the recovery of machine learning models from DRAM under controlled thermal conditions~\cite{won2021deepfreeze}.

Cold boot methods have since been enhanced with cryptanalytic techniques, combining traditional key recovery with quantum-assisted algorithms to improve efficiency~\cite{banegas2023recovering}. Expanding this line of inquiry, researchers have also identified weaknesses in post-quantum cryptographic primitives such as LUOV and AES under cold boot scenarios~\cite{villanueva2020cold, zimerman2021recovering}. The practicality of these attacks remains a concern in modern systems, as shown by ongoing assessments of DRAM remanence and cryptographic leakage~\cite{wetzels2014hidden}. These studies collectively underscore the enduring threat posed by CBA, particularly in cases involving physical access to computing devices. Addressing this vulnerability requires robust encryption key management strategies and hardware-level protections that mitigate data remanence risks.

\subsubsection{\textbf{Peripheral DMA Attacks (PDA)}}

Peripheral DMA attacks take advantage of weaknesses in Direct Memory Access (DMA), which lets peripherals access system memory directly, bypassing the CPU. This design improves performance by offloading data transfer from the processor, but it also creates serious security risks. These attacks typically exploit weaknesses in memory protection or Input-Output Memory Management Unit (IOMMU) configurations to extract sensitive data or compromise system integrity through side-channel techniques~\cite{markettos2019thunderclap}. A foundational study by \textit{Markettos et al.}~\cite{markettos2019thunderclap} revealed critical flaws in \textit{Thunderbolt}. DMA access model, showing how malicious peripherals could bypass IOMMU protections to access arbitrary regions of system memory. Building on this, \textit{Frisk et al.}~\cite{frisk2016direct} demonstrated how attackers could use specially crafted hardware to perform Input-Output Memory Management Unit (IOMMU) bypass and memory extraction. Tools such as \textit{PCILeech} have further operationalized these techniques, enabling full memory acquisition via PCIe interfaces and exposing the practical risk of DMA-based exploitation in modern systems~\cite{frisk18pcileech}.

Research has since expanded to cover more complex deployment scenarios.\textit{Matsuo et al.}~\cite{matsuo2024smmpack} examined the DMA attack surface in multi-tenant environments, where shared access to DMA-capable resources exposed cross-tenant data leakage risks. Similarly, \textit{Butterworth et al.}~\cite{butterworth2013bios} showed that DMA can be used to target System Management Mode (SMM) memory, allowing attackers to access privileged regions without triggering alerts. More recent threats include covert data exfiltration techniques, where PCIe signaling patterns are manipulated to create undetectable communication channels~\cite{tan2021invisible}. Extending these risks to trusted computing, \textit{Van Bulck et al.}~\cite{van2017sgx} demonstrated that DMA can be used to bypass memory isolation in secure enclaves, exposing data from trusted execution environments (TEE) such as Intel SGX.
\subsection{Resource Contention Attacks (RCA)}

Resource Contention Attacks exploit vulnerabilities in shared hardware resources by manipulating contention-induced timing variations or access patterns to extract sensitive data, disrupt operations, or establish covert communication channels~\cite{zhang2012cross}. With the increasing reliance on shared resources in cloud computing, virtualization and high-performance systems. RCA has become a significant threat to system security and performance. RCA can be grouped into three subcategories: cross-VM shared resource attacks (CVSRA), bus and network congestion attacks (BNCA), and GPU memory attacks (GMA). These attacks target resources such as caches, memory buses, network interconnects and GPU, creating measurable side-channel effects and bypassing isolation mechanisms. 

Notably, CVSRA exploits shared caches and interconnects to extract sensitive information across virtual machines~\cite{evtyushkin2016jump}. BNCA, which includes attacks on memory buses, PCIe interconnects and network-on-chip (NoC) architectures, leverages timing-based side channels and covert communication techniques to bypass isolation~\cite{ali2021connoc, chen2024bridge}. In parallel, GMA targets shared GPU memory in multi-tenant environments to facilitate information leakage and covert channels~\cite{dutta2021leaky, chen2024bridge}. Table~\ref{tab:comparison-table-rca} presents the comparison of existing literature that discussed resource contention attacks.

% Please add the following required packages to your document preamble:
% \usepackage{multirow}
\begin{table*}[!ht]
\centering
\caption{Comparison of Resource Contention-Based Memory Side-Channel Attacks}
\label{tab:comparison-table-rca}
\resizebox{\textwidth}{!}{
\begin{tabular}{|c|c|c|c|c|c|c|c|c|c|c|}
\hline
\multirow{2}{*}{\textbf{Literature (Year)}} &
\multirow{2}{*}{\begin{tabular}[c]{@{}c@{}}\textbf{Attack}\\\textbf{Type}\end{tabular}} &
\multirow{2}{*}{\textbf{Methodology}} &
\multicolumn{4}{c|}{\textbf{Platform}} &
\multirow{2}{*}{\textbf{Target}} &
\multirow{2}{*}{\textbf{Impact}} &
\multicolumn{2}{c|}{\textbf{Mitigation}} \\ \cline{4-7} \cline{10-11}
& & & \textbf{Intel} & \textbf{AMD} & \textbf{ARM} & \textbf{Cross} & & & \textbf{HW} & \textbf{SW} \\ \hline\hline

Guo et al. (2022) &
CVSRA &
Prefetch Cache Probing &
\cm &  % Intel
\xm &  % AMD
\xm &  % ARM
\xm &  % Cross
Private Cache Lines &
Covert Channel Leak &
\xm &  % HW Mitigation
\cm \\ \hline  % SW Mitigation

Irazoqui et al. (2015) &
CVSRA &
Cross-VM LLC Probing &
\cm &  % Intel
\xm &  % AMD
\xm &  % ARM
\xm &  % Cross
L3 Cache &
AES Key Recovery &
\xm &  % HW Mitigation
\cm \\ \hline  % SW Mitigation

Lin et al. (2024) &
CVSRA &
In-App Cache Probing &
\xm &  % Intel
\xm &  % AMD
\cm &  % ARM
\cm &  % Cross
Shared Cache (L2/L3) &
App Behavior Inference &
\xm &  % HW Mitigation
\cm \\ \hline  % SW Mitigation

Moghimi et al. (2019) &
CVSRA &
False Dependency Timing &
\cm &  % Intel
\xm &  % AMD
\xm &  % ARM
\xm &  % Cross
Shared Memory Aliasing &
AES Key Recovery &
\xm &  % HW Mitigation
\cm \\ \hline  % SW Mitigation

Shahzad et al. (2015) &
CVSRA &
Cache Profiling &
\cm &  % Intel
\xm &  % AMD
\xm &  % ARM
\cm &  % Cross
Shared CPU Cache &
AES Key Leakage &
\xm &  % HW Mitigation
\xm \\ \hline  % SW Mitigation

Yan et al. (2020) &
CVSRA &
Access Pattern Profiling &
\cm &  % Intel
\cm &  % AMD
\xm &  % ARM
\cm &  % Cross
Last-Level Cache &
DNN Architecture Leak &
\xm &  % HW Mitigation
\cm \\ \hline  % SW Mitigation

Zhang et al. (2012) &
CVSRA &
Prime+Probe Profiling &
\cm &  % Intel
\xm &  % AMD
\xm &  % ARM
\cm &  % Cross
Shared LLC (Cross-VM) &
RSA Key Leakage &
\xm &  % HW Mitigation
\cm \\ \hline  % SW Mitigation

Ristenpart et al. (2009) &
CVSRA &
VM Co-Location Monitoring &
\cm &  % Intel
\xm &  % AMD
\xm &  % ARM
\cm &  % Cross
Shared Physical Hardware &
Co-residency Detection &
\xm &  % HW Mitigation
\cm \\ \hline  % SW Mitigation

Dai et al. (2022) &
BNCA &
Mesh Interconnect Profiling &
\cm &  % Intel
\xm &  % AMD
\xm &  % ARM
\xm &  % Cross
Mesh Network Fabric &
Execution Profiling &
\xm &  % HW Mitigation
\cm \\ \hline  % SW Mitigation

Chacon et al. (2021) &
BNCA &
Coherence Traffic Profiling &
\xm &  % Intel
\xm &  % AMD
\xm &  % ARM
\cm &  % Cross
Interposer Coherence Link &
Covert Channel Leak &
\xm &  % HW Mitigation
\cm \\ \hline  % SW Mitigation

Mustafa et al. (2023) &
BNCA &
PDN Coupling Profiling &
\xm &  % Intel
\xm &  % AMD
\xm &  % ARM
\cm &  % Cross
Interposer Power Grid &
Covert Channel Leak &
\xm &  % HW Mitigation
\cm \\ \hline  % SW Mitigation

Side et al. (2022) &
BNCA &
PCIe Contention Profiling &
\cm &  % Intel
\cm &  % AMD
\xm &  % ARM
\cm &  % Cross
PCIe Bus Interface &
Website Fingerprinting &
\xm &  % HW Mitigation
\xm \\ \hline  % SW Mitigation

Paccagnella et al. (2021) &
BNCA &
Ring Interconnect Profiling &
\cm &  % Intel
\xm &  % AMD
\xm &  % ARM
\xm &  % Cross
CPU Ring Interconnect &
Keystroke and Key Leak &
\xm &  % HW Mitigation
\cm \\ \hline  % SW Mitigation

Savino et al. (2024) &
BNCA &
DRAM Conflict Bombing &
\cm &  % Intel
\xm &  % AMD
\xm &  % ARM
\xm &  % Cross
DRAM Row and Bank &
Timing Amplification &
\xm &  % HW Mitigation
\cm \\ \hline  % SW Mitigation

Bechtel et al. (2023) &
BNCA &
Cache Bank Contention &
\xm &  % Intel
\xm &  % AMD
\cm &  % ARM
\xm &  % Cross
Shared Cache Bank &
Performance Degradation &
\xm &  % HW Mitigation
\cm \\ \hline  % SW Mitigation

Tan et al. (2021) &
BNCA &
PCIe Congestion Timing &
\cm &  % Intel
\cm &  % AMD
\xm &  % ARM
\cm &  % Cross
PCIe Link / Switch &
Activity Inference &
\xm &  % HW Mitigation
\cm \\ \hline  % SW Mitigation

Ali et al. (2021) &
BNCA &
NoC Contention Profiling &
\xm &  % Intel
\xm &  % AMD
\xm &  % ARM
\cm &  % Cross
Network-on-Chip (NoC) &
Covert Channel Leak &
\xm &  % HW Mitigation
\cm \\ \hline  % SW Mitigation

Dutta et al. (2021) &
GMA &
Cross Component Contention &
\cm &  % Intel
\xm &  % AMD
\xm &  % ARM
\xm &  % Cross
Shared LLC Interconnect &
Covert Channel Profiling &
\xm &  % HW Mitigation
\cm \\ \hline  % SW Mitigation

Dutta et al. (2023) &
GMA &
GPU Cache Probing &
\xm &  % Intel
\xm &  % AMD
\xm &  % ARM
\cm &  % Cross
Multi GPU L2 Cache &
ML Model Leak &
\xm &  % HW Mitigation
\cm \\ \hline  % SW Mitigation

% Naghibijouybari et al. (2018) &
% GMA &
% GPU Activity Profiling &
% \xm &  % Intel
% \xm &  % AMD
% \xm &  % ARM
% \cm &  % Cross
% Shared GPU Scheduler &
% User Input Leak &
% \xm &  % HW Mitigation
% \cm \\ \hline  % SW Mitigation

Jin et al. (2024) &
GMA &
Interconnect Profiling &
\xm &  % Intel
\xm &  % AMD
\xm &  % ARM
\cm &  % Cross
GPU On-Chip Interconnect &
RSA Key Leakage &
\xm &  % HW Mitigation
\cm \\ \hline  % SW Mitigation

Karimi et al. (2020) &
GMA &
Memory Coalescing Profiling &
\xm &  % Intel
\xm &  % AMD
\xm &  % ARM
\cm &  % Cross
GPU Coalescing Unit &
AES Key Leakage &
\cm &  % HW Mitigation
\cm \\ \hline  % SW Mitigation

Ferguson et al. (2024) &
GMA &
GPU Prime+Probe &
\cm &  % Intel
\xm &  % AMD
\xm &  % ARM
\xm &  % Cross
GPU L3 Cache &
Website Fingerprinting &
\xm &  % HW Mitigation
\cm \\ \hline  % SW Mitigation

Giner et al. (2024) &
GMA &
WebGPU Prime+Probe &
\cm &  % Intel
\cm &  % AMD
\xm &  % ARM
\cm &  % Cross
GPU L2 Cache &
AES Key Leak &
\xm &  % HW Mitigation
\cm \\ \hline  % SW Mitigation

Cronin et al. (2021) &
GMA &
GPU Contention Timing &
\xm &  % Intel
\xm &  % AMD
\cm &  % ARM
\cm &  % Cross
System-Level Cache (SLC) &
Website Fingerprinting &
\xm &  % HW Mitigation
\cm \\ \hline  % SW Mitigation

Dutta et al. (2021) &
GMA &
Cross-Component Contention &
\cm &  % Intel
\xm &  % AMD
\xm &  % ARM
\cm &  % Cross
LLC + Ring Bus &
Covert Channel Leak &
\xm &  % HW Mitigation
\cm \\ \hline  % SW Mitigation

Zhang et al. (2024) &
GMA &
GPU Interconnect Contention &
\xm &  % Intel
\xm &  % AMD
\xm &  % ARM
\cm &  % Cross
Multi-GPU Link Fabric &
Covert Channel Leak &
\xm &  % HW Mitigation
\cm \\ \hline  % SW Mitigation

\end{tabular}
}
\end{table*}

\subsubsection{\textbf{Cross-VM Shared Resource Attacks (CVSRA)}}

Cross-VM shared resource attacks exploit shared microarchitectural components such as caches, buses, and memory hierarchies to violate isolation boundaries between virtual machines (VMs) in multi-tenant environments. In these scenarios, attackers share the same physical hardware as their victim. They can observe subtle contention patterns or timing differences to infer sensitive data. This makes virtualization boundaries less secure. The concern has grown significantly with the rise of public cloud infrastructure and shared resource models~\cite{cimato2016key}. Early work by \textit{Ristenpart et al.}~\cite{ristenpart2009hey} laid the foundation for this attack surface. In one study, they demonstrated how an attacker could achieve co-residency with a target VM in public cloud infrastructure, revealing the feasibility of strategic VM placement attacks. In a subsequent study, they showed how shared microarchitectural resources such as CPU caches could be leveraged to extract cryptographic keys from neighboring VMs, providing one of the first practical demonstrations of cross-VM side-channel leakage~\cite{zhang2012cross}.

Building on this foundation, \textit{Zhang et al.} revealed that shared last-level caches (LLCs) can be exploited using cache probing techniques to extract data across VM boundaries~\cite{irazoqui2015s}. \textit{Chen et al.} later revealed how modern prefetchers can be misused to construct high bandwidth side channels between VMs~\cite{guo2022adversarial}. More recently, \textit{PREFETCHX}~\cite{chen2024prefetchx} introduced a cross-core, cache-agnostic prefetch-based attack that operates without shared memory, significantly widening the scope of VM-targetable leakage paths. In parallel, \textit{Moghimi et al.} presented \textit{MemJam}~\cite{moghimi2019memjam}, a technique that leverages false memory dependencies to induce contention on shared memory buses and leak information across CPU cores in multi-tenant systems. Expanding the threat landscape further, \textit{Litchfield and Shahzad}~\cite{litchfield2016virtualization} reviewed several cache leakage mechanisms specific to virtualized systems. \textit{Sardar et al.} highlighted how memory deduplication and huge page sharing introduce additional attack vectors between co-resident VMs~\cite{anwar2017cross}. \textit{Yan et al.} demonstrated how page table level access patterns are observable without triggering page faults, It can be used to infer sensitive behavior in co-resident VMs and enclave-like environments~\cite{van2017telling}.

\subsubsection{\textbf{Bus and Network Contention Attacks (BNCA)}}

Bus and network contention attacks exploit shared communication pathways such as memory buses, PCIe interconnects and on-chip networks to breach data confidentiality and disrupt system operations. These attacks typically rely on contention-induced timing variations that allow adversaries to infer sensitive information or establish covert communication channels~\cite{naghibijouybari2022microarchitectural}. For example, \textit{Lee et al.}~\cite{lee2022building} targeted the memory address bus to compromise hardware enclaves, exposing critical vulnerabilities in isolation mechanisms. Expanding the attack surface, \textit{Zhao et al.} demonstrated how PCIe congestion can be exploited to extract GPU workload characteristics, highlighting confidentiality risks in cloud GPU systems~\cite{tan2021invisible}.

Timing-based attacks on memory buses also present serious risks to real-time and critical systems. Studies on DRAM timing leakage have shown that even minor discrepancies in access timing can compromise system integrity in domains such as automotive and aerospace~\cite{savino2024multicore}. Additionally, cache bank contention has been shown to disrupt cross-core operations in multicore processors, revealing potential vectors for denial-of-service via resource exhaustion~\cite{bechtel2023cache}.

Building on these vulnerabilities, recent work on system-on-chip (SoC) architectures has demonstrated that contention-based side channels can extend across subsystems. \textit{Dutta et al.}~\cite{dutta2021leaky} showed that interconnect-level contention between CPU and GPU can be leveraged to build high-bandwidth covert channels in integrated architectures, broadening the threat landscape beyond traditional CPU-centric models.

\subsubsection{\textbf{GPU Memory Attacks (GMA)}}
GPU memory attack exploits the shared architecture of GPU in multi-tenant environments to breach data confidentiality and degrade performance. Using contention-based side channels, attackers can infer sensitive information, disrupt co-located workloads or establish covert communication channels~\cite{naghibijouybari2018rendered}. \textit{Dutta et al.}~\cite{dutta2021leaky} demonstrated covert channels in integrated CPU-GPU systems by exploiting shared interconnects and last-level caches (LLCs), revealing isolation vulnerabilities in heterogeneous environments. Similarly, \textit{Zhang et al.}~\cite{zhang2024beyond} exposed contention risks in NVLink interconnects, showing how congestion in GPU backbones can leak workload behavior in distributed systems. \textit{Zhou et al.}~\cite{naghibijouybari2018rendered} further explored this attack surface by targeting GPU scheduling behavior to infer workload characteristics in NVIDIA devices, enabling sensitive data extraction through timing-based side channels. Extending beyond bare-metal systems, \textit{Jin et al.}~\cite{jin2024ghost} revealed covert channels on virtualized GPU, highlighting architectural weaknesses in cloud environments. In parallel, \textit{Dutta et al.}~\cite{dutta2023spy} demonstrated side channels affecting both performance and confidentiality through contention on shared L2 caches in cloud-based GPU platforms.

As awareness of these vulnerabilities has grown, researchers have proposed mitigation strategies. \textit{Karimi et al.}~\cite{karimi2020hardware} introduced hybrid hardware/software defenses designed to reduce timing leakage without compromising performance. More recently, attention has shifted towards web-based attack surfaces. \textit{Ferguson et al.} demonstrated how WebGPU APIs can be used for high-precision website fingerprinting via cache behavior~\cite{ferguson2024webgpu}, while \textit{Giner et al.}~\cite{giner2024generic} developed automated GPU cache attack primitives capable of launching drive-by attacks with minimal setup. Complementary efforts by \textit{Cronin et al.}~\cite{cronin2021exploration} explored system-level cache-based website fingerprinting using GPU and ARM cache channels, and \textit{Side et al.}~\cite{side2022lockeddown} showed that PCIe bus contention can enable data leakage in both traditional and virtualized environments.
\section{SCAM Mitigation} \label{sec:lesson}

% \subsection{Defense}
Defending against SCAM requires a layered approach that combines software-level protections with hardware-aware defenses. At the software level, techniques such as cryptographic hardening, memory access reshaping, resource isolation and real-time monitoring help reduce exposure by limiting adversarial access to sensitive memory operations~\cite{van2020microarchitectural, srivastava2024scar}. These methods enforce secure execution environments and incorporate runtime anomaly detection to identify malicious behavior during execution~\cite{lou2021survey}. However, such techniques often come with performance overheads or limited coverage.

To address these limitations, countermeasures that directly target microarchitectural leakage channels at the hardware level have emerged. Approaches such as cache randomization, speculative path isolation and prefetch noise addition are designed to disrupt predictable access patterns exploited by attackers~\cite{wang2024safe, qureshi2018ceaser, zhang2024timing}. More advanced defenses include dynamic cache allocation, memory access obfuscation and neural leak detection mechanisms, which introduce greater unpredictability and reduce the effectiveness of attack models~\cite{lyu2018survey}. Additionally, efforts to fix speculative execution leaks, enforce strict control over shared resources, and apply hardware-based transformations offer strong protection with minimal software overhead~\cite{kocher2020spectre}.

Table~\ref{tab:mitigation}  presents different mitigation strategies for specific attack categories. It also specifies the level at which each mitigation should be implemented—whether in hardware, software, or through a hybrid approach combining both \begin{table*}[!ht]
\caption{Mitigation of Different Side Channel Attacks in Memory}
\label{tab:mitigation}
\resizebox{\textwidth}{!}{
\begin{tabular}{|c|c|c|c|c|}
\hline
\textbf{Defense   Technique} & \textbf{Attack Category} & \textbf{Mitigation   Mechanism} & \textbf{Implementation   Level} & \textbf{References} \\ \hline\hline

Cache Partitioning & TBA & Resource Isolation & Hardware & Liu et al. (2020) \\ \hline

Non-Monopolizable Caches & TBA & Simplify contention & Software & Leonid Domnitser et al. (2012) \\ \hline

Deep Learning Detection & TBA & Neural leak detection & Software & Hodong Kim et al. (2023) \\ \hline

Timing Obfuscation & TBA & Obfuscate GPU timing & Software & Karimi et al. (2020) \\ \hline

Timing Repair & TBA & Fix leaks with transformations & Software & Wu et al. (2018) \\ \hline

Timing Framework & TBA & Detect and repair leaks & Algorithmic & Roy et al. (2022) \\ \hline

Constant-Time Primitives & TBA & Fix operation timings & Algorithmic & Montgomery et al. (1987) \\ \hline

Constant-Time RSA & TBA & Uniform RSA timing & Algorithmic & Joye and Yen (2002) \\ \hline

Real-Time Detection & TBA & Spot anomalies instantly & Hardware & Wang et al. (2018) \\ \hline

Trace Analysis & TBA & Minimal-leak detection & Algorithmic & Bronchain et al. (2021) \\ \hline

Timer-Free L3 Cache & TBA & Protect shared resources & Hardware & Craig Disselkoen et al. (2017) \\ \hline

Timing Obfuscation & TBA & Hide memory timing patterns & Algorithmic & Peter W Deutsch et al. (2022) \\ \hline

Cache Isolation Policies & TBA, APA & Isolate speculative paths & Hybrid & Kiriansky et al. (2018) \\
\hline

Transactional Memory & TBA, APA & Isolate transaction states & Hardware & Gruss et al. (2016) \\ 
\hline

Prefetch Noise Injection & TBA, APA & Prefetch noise addition & Software & Luyi Li et al. (2024) \\ \hline

Cache Partitioning & TBA, APA & Dynamic cache allocation & Hybrid & Fiore et al. (2020) \\ \hline

Cache Remapping & TBA, APA & cache reconfiguration & Software & Wei Song et al. (2022) \\ \hline

Vulnerability Analysis & TBA, APA & Automate detection & Software & M Mehdi Kholoosi et al. (2023) \\ \hline

Sensitive Data Protection & TBA, APA & Encrypt and isolate memory & Software & Palit et al. (2019) \\ \hline

Homomorphic Encryption & TBA, APA & Secure computations & Algorithmic & Gentry et al. (2009) \\ \hline

Event Monitoring & TBA, APA & Track system deviations & Hardware & Wu et al. (2022) \\ \hline

Deep Learning & TBA, APA & Use AI to find leaks & Algorithmic & Chiappetta et al. (2016) \\ \hline

Memory Encryption & TBA, APA, FIA & Re-key and mask data & Algorithmic & Xiong et al. (2020) \\ \hline

Dynamic Cache Allocation & TBA, RCA & Cache randomization & Hardware & Werner et al. (2021) \\ \hline

Timing Obfuscation & APA & Obfuscate memory access & Software & Peter W Deutsch et al. (2022) \\ \hline

Privilege Restriction & APA & Restrict sensitive data & Software & Xiaowan Dong et al. (2018) \\ \hline

Obfuscation Engines & APA & Diversify application code & Software & Adil Ahmad et al. (2019) \\ \hline

Cache Analysis & APA & Find leaks early & Algorithmic & Doychev et al. (2015) \\ \hline

Non-Monopolizable Caches & APA & Mitigate contention & Hardware & Leonid Domnitser et al. (2012) \\ \hline

Scatter-Gather Defense & APA, TBA & Prevent timing leaks & Algorithmic & Yarom et al. (2017) \\ \hline

Page Table Randomization & APA, RCA & Randomize memory layouts & Software & Raoul Strackx et al. (2017) \\ \hline

Memory Tagging & APA, RCA & Break speculative tagging & Software & Juhee Kim et al. (2024) \\ \hline

SGX Randomization & APA, RCA & Randomize memory mappings & Software & Brasser et al. (2017) \\ \hline

Isolation Mechanisms & APA, RCA & Partition shared memory & Hardware & Strackx et al. (2023) \\ \hline

Memory Obfuscation & APA, RCA & Hide memory access patterns & Algorithmic & Jiang et al. (2020) \\ \hline

Counter Detection & APA, RCA & Detect attacks with counters & Hardware & Carnà et al. (2023) \\ \hline

Memory Obfuscation & SBA & Hide memory patterns & Software & Jiang et al. (2020) \\ \hline

Encrypted Memory & SBA & Mask and encrypt data & Hardware & Moëllic et al. (2021) \\ \hline

GPU Timing Protection & SBA & Obfuscate GPU timing & Hybrid & Karimi et al. (2020) \\ \hline

Signal Control & SBA & Lower EM and power signals & Hardware & Das et al. (2023) \\ \hline

In-Memory Security & SBA & Protect in-memory data & Hardware & Das et al. (2022) \\ \hline

Row Refresh Mechanism & FIA & Control row activations & Hardware & Kim et al. (2014) \\ \hline

Rowhammer Defense & FIA & Block Rowhammer attacks & Algorithmic & Derya et al. (2024) \\ \hline

Fault Defense Intro & FIA & Analyze and defend faults & Conceptual & Kim and Ha (2015) \\ \hline

Self-Reducibility & FIA & Use faults smartly & Algorithmic & Erata et al. (2024) \\ \hline

Software Evaluation & FIA & Test fault defenses & Software & Moro et al. (2014) \\ \hline

Ring Oscillator & FIA, APA & Block faults and channels & Hardware & Yao et al. (2021) \\ \hline

Core Pinning & RCA & Dedicate cores & Hardware & Read Sprabery et al. (2018) \\ \hline

Cache Protections & RCA & Prevent eviction with locks & Software & Taesoo Kim et al. (2012) \\ \hline

Cache Randomization & RCA & Randomize cache states & Hardware & Wang et al. (2020) \\ \hline

Cache Timing Detection & RCA & Analyze timing thresholds & Algorithmic & Sangeetha et al. (2021) \\ \hline

Cache Partitioning & RCA & Dynamic cache allocation & Hardware & Fiore et al. (2020) \\ \hline

Cache Remapping & RCA & Cache reconfiguration & Hardware & Wei Song et al. (2022) \\ \hline

Core Pinning & RCA & Use dedicated cores & Software & Read Sprabery et al. (2018) \\ \hline

Bus Monitoring & RCA & Analyze bus patterns & Hardware & Raj et al. (2020) \\ \hline

% \end{tabular}
% }
% \end{table*}

% \begin{table*}[!ht]
% \ContinuedFloat 
% \caption{Mitigation of Different Side-Channel Attacks}
% \label{tab:mitigation}
% \resizebox{\textwidth}{!}{
% \begin{tabular}{|c|c|c|c|c|}
% \hline
% \textbf{Defense   Technique} & \textbf{Attack Category} & \textbf{Mitigation   Mechanism} & \textbf{Implementation   Level} & \textbf{References} \\ \hline
% \hline

\end{tabular}
}
\end{table*}

\subsection{Hardware Level Mitigations (HLM)}

Hardware-level mitigations form the root layer of protection against SCAM by addressing vulnerabilities intrinsic to physical and microarchitectural components. These defenses are designed to counter adversaries who exploit timing variations, access patterns, or speculative execution flaws to extract sensitive information. As advanced techniques like \textit{Prime+Probe} and \textit{Rowhammer} continue to evolve, hardware-based solutions have become indispensable for safeguarding modern computing systems~\cite{liu2015last, mutlu2019rowhammer}.

A core focus of hardware mitigation is cache architecture. By redesigning how caches manage memory, systems can reduce timing discrepancies that side-channel attacks rely on. Partitioning techniques such as cache coloring enforce process isolation at the hardware level. \textit{Liu et al.}~\cite{liu2016catalyst} proposed dedicating specific cache sets to individual processes to prevent resource contention and reduce leakage in shared environments. \textit{Werner et al.}~\cite{werner2019scattercache} introduced dynamic cache set randomization, making it difficult for attackers to predict access patterns. To defend against speculative execution attacks, \textit{Kiriansky et al.}~\cite{kiriansky2018dawg} implemented strict cache isolation policies, while \textit{Gruss et al.}~\cite{gruss2017strong} utilized hardware transactional memory to protect sensitive code paths by isolating execution states. Additionally, \textit{Chen et al.}~\cite{chen2018racing} proposed dynamic cache flushing to periodically clear shared caches, thereby mitigating speculative traces and residual data leakage.

Beyond caches, memory-level technologies have advanced to prevent fault and access-based attacks. \textit{Rowhammer} resistant DRAM designs like target row refresh (TRR), \textit{PROTRR} monitor and regulate activation frequencies to prevent disturbance errors~\cite{frigo2020trrespass, marazzi2022protrr}. Error-Correcting Code (ECC) memory adds fault resilience by automatically detecting and correcting bit flips in adversarial scenarios~\cite{kim2023kill}. \textit{Kim et al.}~\cite{kim2014flipping} introduced DRAM randomization, dynamically changing physical memory layouts to obscure timing-based patterns. \textit{Dessouky et al.}~\cite{dessouky2020hybcache} combined static and dynamic cache configurations to balance performance and security in trusted platforms. Similarly, \textit{Godfrey et al.}~\cite{godfrey2014preventing} proposed tenant isolation mechanisms in cloud environments to prevent cross-VM cache-based leakage.

At the processor level, secure hardware design mitigates microarchitectural risk such as speculative leakage. Speculation barriers like \textit{SafeSpec}~\cite{khasawneh2019safespec} and \textit{PoisonIvy}~\cite{ lehman2016poisonivy} prevent transient instructions from leaving observable states. \textit{Yan et al.} developed \textit{InvisiSpec}~\cite{yan2018invisispec} to suppress speculative execution traces from becoming accessible to attackers. Secure enclave extensions, such as enhanced SGX designs, strengthen defenses against controlled-channel attacks by hiding memory access patterns~\cite{shih2017t}. \textit{Van Bulck et al.}~\cite{van2018foreshadow} introduced architectural separations between privileged and non-privileged modes to limit leakage across protection domains.

Despite trade-offs in cost and performance, hardware-level mitigations provide a robust root of trust against SCAM. By embedding protection directly into the architecture, they ensure long-term resilience against evolving consumer and cloud system threats.

\subsection{\textbf{Software Level Mitigations (SLM)}} Software level mitigations play a pivotal role in defending against SCAM by addressing vulnerabilities at the operating system, runtime, and application levels~\cite{zhou2016software}.
Unlike hardware-based solutions, these defenses are flexible, cost-effective, and deployable on existing platforms without requiring physical changes~\cite{brasser2017dr}.
By modifying memory access patterns, obfuscating execution behavior, and introducing randomness, software-level defenses offer an essential line of protection against evolving threats~\cite{crane2015thwarting}.

A primary approach involves retrofitting existing systems to reduce leakage opportunities in multi-tenant environments. For example, \textit{Zhou et al.}~\cite{zhou2016software} proposed mechanisms to reduce cache interference, combining speculative execution hardening and obfuscation to build layered defenses. At the OS level, \textit{Zhang et al.}~\cite{zhang2013duppel} introduced artificial noise into memory access patterns to disrupt cache-based side channels in virtualized systems, while \textit{Orenbach et al.}~\cite{orenbach2017eleos} developed exitless system calls for SGX, minimizing enclave exits and reducing attack surfaces. Runtime memory randomization further obscures predictable mappings during program execution~\cite{chen2016remix}. Another important strategy targets speculative execution and cache timing attacks. \textit{Khasawneh et al.} introduced \textit{SafeSpec}~\cite{khasawneh2019safespec}, which transparently revokes speculative memory operations before they can leak information. Complementary techniques include randomized cache line replacement policies to hinder timing inference~\cite{wang2020mitigating}, and privilege separation mechanisms to eliminate high-privilege leakage channels~\cite{dong2018shielding}. \textit{Doychev et al.}~\cite{doychev2017rigorous} evaluated a range of such defenses to provide insight into their practical impact.

Page-level defenses have also proven effective. Techniques like page table randomization~\cite{strackx2017heisenberg} and memory layout diversification in SGX~\cite{ahmad2019obfuscuro} limit deterministic leakage paths. Code diversification and dynamic software diversity mechanisms increase attacker uncertainty~\cite{crane2015thwarting}, while automated transformations have been proposed to eliminate timing-based leakage systematically~\cite{li2020dlfix}. In addition, \textit{Li et al.}~\cite{li2024p} introduced intelligent prefetching to inject noise into memory access patterns. Further protections focus on securing memory-resident sensitive data. Encryption and isolation strategies help prevent leakage in shared systems~\cite{palit2019mitigating}, and randomizing data placement within enclaves blocks deterministic inference~\cite{brasser2017dr}. System-level defenses like cache line locking~\cite{kim2012stealthmem} and page table isolation~\cite{tan2023ptstore} offer strong protection against \textit{Meltdown}~\cite{lipp2020meltdown} attack and residual attacks.

Despite potential trade-offs in performance and system complexity, software-level defenses are indispensable. Their adaptability makes them a crucial complement to hardware-based mechanisms in building resilient platforms against SCAM.

\subsection{\textbf{Cryptographic and Algorithmic Mitigations (CAM)}} Cryptographic and algorithmic strategies offer a specialized cryptographic shield by redesigning protocols and computations to minimize exploitable leakage~\cite{kocher1996timing, wichelmann2023cipherfix}. These techniques help protect sensitive operations by hiding timing, power, and memory access patterns that attackers could use to steal information.

A central strategy involves adopting robust cryptographic designs that conceal key-dependent behavior. Techniques such as hardware transactional memory isolate critical code paths and suppress intermediate leakage, while execution path obfuscation introduces noise into control flow, making timing-based analysis less reliable~\cite{rane2015raccoon}. Algorithmic masking and blinding add randomized elements to intermediate values, especially in elliptic curve cryptography, to disrupt deterministic execution traces~\cite{goubin2002refined}. These measures help ensure that sensitive data does not correlate with observable system behavior. Beyond cryptographic routines, dynamic memory defense against obscure exploitable patterns. Memory polymorphism, as introduced by \textit{Jiang et al.}, randomizes memory layout during runtime to prevent deterministic memory tracing~\cite{jiang2020mempoline}. When combined with polymorphic code encryption, these approaches protect against ciphertext leakage and ensure integrity even under partial memory disclosure~\cite{morel2023code}. Homomorphic encryption offers an even more powerful abstraction by enabling computations directly on encrypted data, supporting secure processing in multi-tenant or untrusted environments~\cite{gentry2009fully}.

To counter timing-based attacks, constant-time algorithms such as the Montgomery ladder are designed to run with uniform control flow and memory access regardless of input~\cite{joye2002montgomery}. These techniques are further supported by frameworks like \textit{CipherFix}~\cite{wichelmann2023cipherfix}, which enforce timing independence across cryptographic functions. \textit{Wu et al.}~\cite{wu2018eliminating} proposed a systematic software transformation pipeline to detect and eliminate timing-based side-channel vulnerabilities in real-world applications. Additional layers of defense include memory encryption and authentication protocols. Leakage-resilient protocol designs eliminate timing variability in key exchange mechanisms~\cite{dziembowski2008leakage}. Techniques like MEAS (Memory Encryption with Authentication and Splitting) use re-keying, authentication trees, and masking to protect data during memory operations~\cite{unterluggauer2019meas}. Threshold cryptography further distributes risk by splitting secret keys into shares, reducing the impact of single-point leakage. Optimized cryptographic implementations have also been hardened to resist power analysis by flattening execution patterns and minimizing energy-related signatures~\cite{tsoupidi2023securing}.

As side-channel attack methodologies continue to evolve, cryptographic and algorithmic mitigations remain essential in layered defense models. These approaches must continue to innovate to provide strong, adaptable protection across diverse computing environments.

\subsection{\textbf{Isolation and Resource Partitioning (IRP)}} Isolation and partitioning techniques act as a structural barrier against SCAM by enforcing resource separation and limiting adversarial interference.~\cite{varadarajan2014scheduler}. These strategies are especially critical in multi-tenant and virtualized environments, where attackers may exploit shared resource contention to bypass isolation mechanisms.

A prominent defense mechanism involves cache isolation and partitioning, which restricts how adversarial processes interact with shared cache hierarchies~\cite{domnitser2012non}. Low complexity cache designs prevent cache monopolization by isolating cache sets or ways, thereby reducing contention in shared environments~\cite{wang2007new}. Technologies like Intel CAT enable dynamic partitioning of last-level caches (LLCs), minimizing interference between trusted and untrusted processes~\cite{liu2016catalyst}. \textit{Song et al.}~\cite{song2022remapped} proposed remapped cache layouts to dynamically assign cache regions to different workloads, breaking predictable access patterns while preserving performance. Set-based partitioning further enhances isolation by assigning specific cache sets to individual threads or applications~\cite{dessouky2021chunked}. 

Complementing these hardware controls, scheduler-based techniques offer temporal and spatial process isolation. CPU core pinning dedicates processor cores to sensitive tasks, reducing the risk of cross-thread leakage in shared environments~\cite{sprabery2018scheduling}. Advanced task scheduling algorithms further enhance resilience by ensuring that adversarial and critical workloads do not run concurrently on shared resources~\cite{varadarajan2014scheduler}. When integrated with cache allocation policies, these scheduling strategies allow secure tasks to be mapped to isolated resources, balancing throughput with security~\cite{sprabery2018scheduling}. Beyond CPU and cache-level protections, memory isolation techniques address vulnerabilities in access patterns and physical memory exposure. Memory access isolation prevents observable patterns that could reveal enclave-level information~\cite{oleksenko2018varys}. Obfuscation strategies, such as randomized memory access sequences, further disrupt timing-based inference attacks~\cite{deutsch2022dagguise}. For stronger confidentiality guarantees, memory encryption ensures that virtual machine (VM) memory remains inaccessible even under physical compromise~\cite{geng2024new}.

Taken together, these isolation and resource partitioning strategies form a robust defense against memory side-channel threats. By integrating separation mechanisms at the cache, CPU, and memory levels, IRP approaches mitigate cross-domain interference while maintaining performance and scalability in modern multi-tenant systems.

\subsection{\textbf{Detection and Monitoring (DM)}} 
Detection and monitoring techniques form a critical line of defense against SCAM by integrating static code analysis, real-time system monitoring, machine learning-based anomaly detection, and hardware-level event tracking. These approaches operate across both software and hardware layers, enabling timely detection, response, and adaptation to evolving threat landscapes.

Static detection mechanisms play a crucial role during the development phase by identifying vulnerabilities before deployment. Tools like \textit{CacheAudit}~\cite{doychev2015cacheaudit}, developed by \textit{Doychev et al.}, analyze cache usage to detect potential leakage paths during code development, allowing developers to apply preemptive countermeasures. Similarly, symbolic execution frameworks such as \textit{CaSym}~\cite{brotzman2019casym} trace execution paths to identify sensitive code blocks susceptible to side-channel exposure. Dynamic detection techniques enhance system resilience by monitoring behavior at runtime to detect ongoing attacks. Hardware Performance Counters (HPC) can be leveraged to identify anomalous memory or cache access patterns indicative of side-channel activity~\cite{chiappetta2016real}. \textit{Wang et al.}~\cite{wang2017cached} demonstrated timing channel detection in live environments, while \textit{Wu et al.}’s \textit{PREDATOR}~\cite{wu2022predator} system introduced threshold-based event monitoring to trigger automated mitigations in response to suspicious performance deviations.

Machine learning-based approaches complement static and dynamic defenses by identifying complex behavioral patterns that traditional methods might overlook. Deep learning models trained on execution traces can detect subtle timing or memory anomalies. \textit{Kim et al.}~\cite{kim2023deep} proposed a neural network-based detection technique that accurately classifies memory access behaviors, even under obfuscation or adaptive attack conditions. Hardware-assisted mechanisms offer low-level monitoring capabilities to detect and respond to attacks with minimal performance overhead. \textit{Wang et al.}~\cite{wang2020scarf} utilized embedded hardware features to enable real-time, high-fidelity attack detection. \textit{Raj et al.}~\cite{raj2017keep} introduced memory bus monitoring and cache obfuscation to reduce leakage visibility. Additional techniques, such as interrupt-driven alerts, can halt execution when attack signatures are detected, while process throttling dynamically limits the resource access of suspicious workloads.

Detection and monitoring techniques work in layers to build an adaptive defense framework. By combining static analysis, runtime anomaly tracking, intelligent behavioral models, and hardware-integrated safeguards, modern systems can proactively identify and respond to memory side-channel threats. Recent developments demonstrate this integration in practice: \textit{Carnà et al.}~\cite{carna2023fight} leverage hardware-assisted detection for rapid response to cache-based attacks, while \textit{Sangeetha et al.}~\cite{sangeetha2021optimistic} apply predictive runtime techniques to reduce detection latency and improve coverage. Such multifaceted approaches highlight the importance of coordination across detection layers to ensure resilient and responsive protection in complex computing environments.

% \subsection{Performance Impact of Countermeasures}

\section{Conclusion and Future Work} \label{sec:conclusion}

\subsection{Related Work}

Side-channel attacks on memory have been the focus of extensive research, with numerous surveys examining various aspects of these threats. However, many of these works are platform-specific or limited to particular attack categories, leaving space for a more unified perspective that captures emerging vulnerabilities alongside traditional ones.

The recent contributions of \textit{Pulkit et al.}~\cite{pulkit2023survey} provide a structured taxonomy of cache side-channel attacks, examining microarchitectural contexts and suggesting mitigation strategies. Similarly, \textit{Zhang et al.}~\cite{zhang2014cache} and \textit{Lou et al.}~\cite{lou2021survey} explore cache timing attacks in multi-tenant environments, offering detailed classification frameworks. Although these works are foundational, they primarily address timing-based cache attacks and do not incorporate newer vectors such as speculative execution or GPU-based threats~\cite{canella2019systematic, gruss2018another, kim2020gpu}.

Other surveys, including \textit{Szefer et al.}~\cite{szefer2019survey}, focus on covert microarchitectural channels but do not extend their scope to DRAM or transient execution vulnerabilities. \textit{Lou et al.}~\cite{lou2021survey} also examine cryptographic side-channel threats, yet their analysis remains limited to specific contexts and lacks system-level integration. More recently, studies by \textit{Lungu et al.}~\cite{lungu2024gpu} and \textit{Wang et al.}~\cite{wang2023side} investigate side channels in GPU memory and neural network architectures, signaling the expansion of attack surfaces in modern hardware. However, these efforts often address isolated attack vectors without connecting them to broader threat taxonomies.

To address these gaps, this survey presents a unified taxonomy that bridges conventional and emerging memory side-channel attacks. Strategically linking attack methodologies to corresponding mitigations provides a cohesive framework for understanding and countering side-channel attack mechanisms. Moreover, this work synthesizes fragmented findings into an organized narrative, tracing the evolution of attack surfaces while categorizing defenses across diverse architectures and computing environments.

\subsection{Future Research Directions}

As SCAM continues to evolve, future research must focus on proactive mitigation techniques and deeper exploration of emerging vulnerabilities. One key direction is the development of hardware-level defenses that can effectively counter SCAM without significant performance trade-offs. Existing mitigations, such as cache partitioning and constant-time execution, offer partial solutions but are not universally effective across all SCAM categories. Future architectures must incorporate microarchitectural modifications, enhanced memory encryption techniques and intelligent resource isolation mechanisms to mitigate SCAM at its root cause. Additionally, formal security verification models for hardware design can help identify and eliminate vulnerabilities before they are exploited.

On the software front, compiler-assisted mitigations and OS-level protections present promising research potential. Techniques, like control flow obfuscation, randomized memory layouts, and secure memory access patterns, could help reduce SCAM exposure. However, since adversaries constantly find novel ways to bypass these defenses, researchers need to find better adaptive and machine-learning-driven detection mechanisms. Research on dynamic anomaly detection using AI-driven monitoring of memory behavior could provide real-time countermeasures against SCAM in cloud and multi-tenant environments. 

Lastly, given the increasing adoption of trusted execution environments (TEE) and confidential computing, future research should explore SCAM-resistant enclave designs and secure hypervisor-level protections. The intersection of SCAM with emerging attack vectors such as transient execution vulnerabilities and speculative execution side channels also warrants further investigation. As adversaries continue refining their techniques, a collaborative effort among academia, industry, and hardware manufacturers will be essential to developing robust, long-term solutions against SCAM.

%%
%% The acknowledgments section is defined using the "acks" environment
%% (and NOT an unnumbered section). This ensures the proper
%% identification of the section in the article metadata, and the
%% consistent spelling of the heading.

%%
%% The next two lines define the bibliography style to be used, and
%% the bibliography file.
\bibliographystyle{ACM-Reference-Format}
\bibliography{references.bib}

% \clearpage
% \input{Sections/appendix}

%%
\end{document}